\documentstyle[11pt,axodraw]{article}

\font\tenrm=cmr10
\font\tenit=cmti10
\font\elevenbf=cmbx10 scaled\magstep 1
\font\elevenrm=cmr10 scaled\magstep 1
\font\elevenit=cmti10 scaled\magstep 1

\def\roughly#1{\mathrel{\raise.3ex\hbox{$#1$\kern-.75em\lower1ex\hbox{$\sim$}}}}
\def\lsim{\roughly<}

\def\Square{{\vbox {\hrule height 0.6pt\hbox{\vrule width 0.6pt\hskip 3pt
        \vbox{\vskip 6pt}\hskip 3pt \vrule width 0.6pt}\hrule height 0.6pt}}}
\def\beq{\begin{equation}}
\def\eeq{\end{equation}}

\renewenvironment{thebibliography}[1]
 { \elevenrm
   \begin{list}{\arabic{enumi}.}
    {\usecounter{enumi} \setlength{\parsep}{0pt}
     \setlength{\itemsep}{3pt} \settowidth{\labelwidth}{#1.}
     \sloppy
    }}{\end{list}}

\begin{document}
\begin{flushright}
FERMILAB-Pub-95/031\\
{\bf MAD/PH/871}\\
UB-HET-95-01\\
UdeM-GPP-TH-95-14\\
March 1995\\
\end{flushright}
\begin{center}{
\vglue 0.6cm
{\elevenbf ANOMALOUS GAUGE BOSON INTERACTIONS
\footnote{Summary of the Working Subgroup on Anomalous Gauge Boson
Interactions of the DPF Long-Range Planning Study,
to be published in ``Electroweak
Symmetry Breaking and Beyond the Standard Model'', eds. T. Barklow,
S. Dawson, H. Haber and J. Siegrist. }
\\ }
\vglue 0.4cm
\baselineskip=13pt
{\tenrm H.~Aihara$^1$\footnote{Co-convener}, T.~Barklow$^2$,
U.~Baur$^{3,4\dagger}$, J.~Busenitz$^5$, S.~Errede$^6$, T.~A.~Fuess$^7$,
T.~Han$^8$, D.~London$^{9\dagger}$, J.~Ohnemus$^8$, R.~Szalapski$^{10}$,
C.~Wendt$^{11}$, and D.~Zeppenfeld$^{11\dagger}$ \\}
\vglue 0.3cm
\baselineskip=12pt
{\tenit $^1$Lawrence Berkeley Laboratory, Berkeley, CA 94720\\
$^2$Stanford Linear Accelerator Center, Stanford, CA 94309\\
$^3$Physics Department, SUNY Buffalo, Buffalo, NY 14260\\
$^4$Physics Department, Florida State University, Tallahassee, FL 32306 \\
$^5$Department of Physics and Astronomy, University of Alabama,
Tuscaloosa, AL 35487\\
$^6$Physics Department, University of Illinois at Urbana -- Champaign,
Urbana, IL 61801\\ $^7$Argonne National Laboratory, Argonne, IL 60439\\
$^8$Physics Department, University of California, Davis, CA 95616\\
$^9$Physics Department, University of Montreal, Canada H3C 3J7\\
$^{10}$Theory Group, KEK, Tsukuba, Ibaraki 305, Japan\\
$^{11}$Physics Department, University of Wisconsin, Madison, WI 53706\\}
\vglue 0.8cm
{\tenrm ABSTRACT}}

\end{center}

\vglue 0.1cm
{\rightskip=3pc
 \leftskip=3pc
 \tenrm\baselineskip=12pt
 \noindent
We discuss the direct measurement of the trilinear vector boson couplings
in present and future collider experiments. The major goals of such
experiments will be the confirmation of the Standard Model (SM)
predictions and the search for signals of new physics. We
review our current theoretical understanding of anomalous trilinear
gauge-boson self interactions. If the energy scale of the new physics is
$\sim 1$ TeV, these low energy anomalous couplings are expected to be
no larger than
${\cal O}(10^{-2})$. Constraints from high precision measurements at
LEP and low energy charged and neutral current processes are critically
reviewed.
\vglue 0.6cm}
{\elevenbf\noindent 1. Introduction}
\vglue 0.2cm
\baselineskip=14pt
\elevenrm
Over the last five years $e^+e^-$ collision experiments at LEP and at
the SLAC
linear collider have beautifully confirmed the predictions of the Standard
Model (SM). At present experiment and theory agree at the 0.1~--~1\% level
in the determination of the vector boson couplings to the various
fermions~\cite{rolandi}, which may rightly be considered a confirmation
of the
gauge boson nature of the $W$ and the~$Z$. Nevertheless the most direct
consequences of the $SU(2)_L\times U(1)_Y$ gauge symmetry, the non-abelian
self-couplings of the $W$, $Z$, and photon, remain poorly measured to date.

A direct measurement of these vector boson couplings is possible in
present and future collider experiments, in particular via pair
production processes like
$e^+e^- \to W^+W^-$, $Z\gamma$ and  $q \bar q \to W^+W^-,\; W\gamma,\;
Z\gamma,\; WZ$. The first and
major goal of such experiments will be a confirmation of the SM
predictions. A precise and direct measurement of the
trilinear and quartic couplings of the electroweak vector bosons
and the demonstration that they agree with the SM would
beautifully corroborate spontaneously broken, non-abelian gauge
theories as the basic theoretical structure describing the fundamental
interactions of nature.

At the same time, such measurements may be used to probe for new physics.
Since the gauge boson self-couplings have not yet been measured with good
precision, it is possible in principle that signals for physics  beyond the
SM will appear in this sector through the discovery of {\elevenit
anomalous} trilinear (or quartic) gauge-boson vertices (TGV's). This
possibility
immediately raises a number of other questions. What are the expected sizes
of such anomalous effects in different models of new physics? Will the new
physics which gives rise to anomalous gauge-boson couplings manifest itself
in other observables and/or other channels? Are there significant
constraints from low-energy measurements? We address these questions in
Section~2.

For the most part, however, we are interested in the accuracy of
various collider experiments for a
direct measurement of the self-interactions of electroweak vector
bosons, so as to evaluate how well the SM predictions can be tested
(Section~3). For simplicity, we
shall restrict ourselves to trilinear vector boson couplings, in particular
the $WWV$, and $Z\gamma V$, $V=\gamma,\, Z$ couplings. Possibilities to
test quartic couplings in collider experiments are discussed in
Ref.~\cite{QUARTIC}.

Analogous to the introduction of arbitrary vector and axial vector
couplings $g_V$ and $g_A$ for the coupling of gauge bosons to
fermions, the measurement of the $WWV$ couplings can be made quantitative by
introducing a more general $WWV$ vertex. For our discussion of experimental
sensitivities in Section~3 we shall use a parameterization in terms of the
phenomenological effective Lagrangian~\cite{HPZH}
\begin{eqnarray} \label{LeffWWV}
i{\cal L}_{eff}^{WWV} & = & g_{WWV}\, \Bigl[ g_1^V\left(
W^{\dagger}_{\mu\nu}W^{\mu}-W^{\dagger\, \mu}W_{\mu\nu}\right)V^{\nu} +
\kappa_V\,  W^{\dagger}_{\mu}W_{\nu}V^{\mu\nu} + \\ \nonumber
& & {\lambda_V\over m_W^2}\,
W^{\dagger}_{\rho\mu}{W^{\mu}}_{\nu}V^{\nu\rho}
+ig_5^V\varepsilon_{\mu\nu\rho\sigma}\left(
(\partial^\rho W^{\dagger\, \mu})W^\nu -
W^{\dagger\, \mu}(\partial^{\rho}W^{\nu}) \right) V^{\sigma} \Bigr]\; .
\end{eqnarray}
Here the overall couplings are defined as $g_{WW\gamma}=e$ and
$g_{WWZ}= e \cot\theta_W$, $W_{\mu\nu}=\partial_\mu W_\nu - \partial_\nu
W_\mu$, and $V_{\mu\nu}=\partial_\mu V_\nu - \partial_\nu V_\mu$. Within
the SM, at tree level, the couplings are given by $g_1^Z = g_1^\gamma =
\kappa_Z = \kappa_\gamma = 1,\; \lambda_Z = \lambda_\gamma = g_5^Z =
g_5^\gamma = 0$. For on-shell photons, $g_1^\gamma=1$ and
$g_5^\gamma=0$ are
fixed by electromagnetic gauge invariance; $g_1^Z$ and $g_5^Z$ may,
however, differ from their SM values. Deviations are given by the
anomalous TGV's
\beq
\label{anomTGVs}
\Delta g_1^Z \equiv (g_1^Z - 1)~,~~~ \Delta \kappa_\gamma \equiv
(\kappa_\gamma-1)~,~~~ \Delta\kappa_Z \equiv (\kappa_Z-1)~,~~~
\lambda_\gamma~,~~~ \lambda_Z~,~~~ g_5^Z ~.
\eeq
As we discuss in Section 2, theoretical arguments suggest that
these anomalous TGV's are at most of ${\cal O}(m_W^2/\Lambda^2)$, where
$\Lambda$ is the scale of new physics
(some are expected to be considerably smaller). Thus, for $\Lambda\sim
1$~TeV, the anomalous TGV's are ${\cal O}(10^{-2})$, which will make their
observation
difficult. Conversely, if large anomalous TGV's are discovered, this
implies that the new physics responsible for them is likely to be found
directly below the TeV scale.

The effective Lagrangian of Eq.~(\ref{LeffWWV}) parameterizes the
most general
Lorentz invariant and $CP$ conserving $WWV$ vertex which can be observed in
processes where the vector bosons couple to effectively massless
fermions. Apart from $g_5^V$, all couplings conserve $C$ and $P$
separately. If $CP$ violating couplings are allowed, three additional
couplings, $g_4^V$, $\tilde\kappa_V$ and $\tilde\lambda_V$, appear in the
effective Lagrangian~\cite{HPZH} and they all vanish in the SM, at tree
level. For simplicity, these couplings are
not considered in this report. Terms
with higher derivatives are equivalent to a dependence of the couplings
on the vector boson momenta and thus merely lead to a form-factor
behaviour of these couplings (see Section~2.3). The $C$ and $P$
conserving terms in ${\cal L}_{eff}^{WW\gamma}$
correspond to the lowest order terms in a multipole expansion of the
$W-$photon interactions, the charge $Q_W$, the magnetic dipole
moment $\mu_W$ and the electric quadrupole moment $q_W$  of the
$W^+$~\cite{AR}:
\begin{eqnarray}\label{EQ:multipole}
Q_W & = & e g_1^\gamma\; , \\
\mu_W & = & {e \over 2m_W}
\left(g_1^\gamma + \kappa_\gamma + \lambda_\gamma\right) \; , \\
q_W & = & -{e \over m_W^2} \left(\kappa_\gamma-\lambda_\gamma\right) \; .
\end{eqnarray}

Analogous to the general $WWV$ vertex it is possible to parameterize
anomalous $Z\gamma V,\, V=\gamma ,Z$ couplings. We shall be interested in
constraints from $Z\gamma$ production processes in Section~3,
{\elevenit i.e.} we may treat the
photon and the $Z$ as being on-shell.  As before we are only considering
$CP$-even couplings.  Let us denote the Feynman rule for the
$V_\mu(P) \to Z_\alpha(q_1)\gamma_\beta(q_2)$ vertex by
$ie\Gamma^{\alpha\beta\mu}_{Z\gamma V}(q_1,q_2,P)$. The most general
such vertex compatible with Lorentz invariance has been discussed in
Ref.~\cite{HPZH} and it
can be parameterized in terms of two free parameters, $h_3^V$ and $h_4^V$,
\begin{equation}\label{EQ:ZgammaV}
\Gamma^{\alpha\beta\mu}_{Z\gamma V}(q_1,q_2,P) = {P^2-m_V^2 \over m_Z^2}
\Bigl[ h_3^V \varepsilon^{\mu\alpha\beta\rho}q_{2\rho} +
{h_4^V \over m_Z^2} P^\alpha
\varepsilon^{\mu\beta\rho\sigma}P_{\rho} q_{2\sigma} \Bigr]\; .
\end{equation}
Within the SM, at tree level, $h_3^V=h_4^V=0$.
If $CP$ violating couplings are allowed, two additional
couplings, $h_1^V$ and $h_2^V$, appear in the
effective Lagrangian~\cite{HPZH} which also vanish in the SM, at tree
level. For simplicity, these couplings are not considered here.
The overall factor $P^2-m_V^2$ in Eq.~(\ref{EQ:ZgammaV}) is implied by
Bose symmetry for on-shell $V$ and/or by gauge invariance for
$V=\gamma$. These additional factors
indicate that anomalous $Z\gamma V$ couplings can only arise from higher
dimensional operators than the $WWV$ couplings and hence their effects
should be suppressed in any scenario of new physics beyond the SM.

In Section~3
present measurements from $p\bar p$ and $e^+e^-$ collider experiments
are summarized.
In addition the sensitivity of future Tevatron, LHC, LEP~II and NLC
experiments is analysed in detail. Our conclusions are presented
in Section~4.

\vglue 0.3cm
{\elevenbf\noindent 2. Theoretical Background}
\vglue 0.2cm
{\elevenit\noindent 2.1 Effective Lagrangians: General Considerations}
\vglue 0.2cm

In this Section we discuss theoretical ideas which lead to anomalous gauge
boson self-interactions, and analyze constraints from low energy and
high precision measurements.
In the absence of a specific model of new physics, effective Lagrangian
techniques are extremely useful. An effective Lagrangian~\cite{efflag}
parameterizes, in a model-independent way, the low-energy effects of the new
physics to be found at higher energies. It is only necessary to specify the
particle content and the symmetries of the low-energy theory. Although
effective Lagrangians contain an infinite number of terms, they are
organized in powers of $1/\Lambda$, where $\Lambda$ is the scale of new
physics. Thus, at energies which are much smaller than $\Lambda$, only the
first few terms of the effective Lagrangian are important.

The Fermi theory of the weak interactions is perhaps the best-known example
of an effective Lagrangian. Within the SM, the charged-current interaction
between two fermions is described by the exchange of a $W$-boson:
\beq
\label{SMCC}
{g^2\over 8}\, {\overline\Psi}\gamma_\mu (1-\gamma_5) \Psi \,
{1\over q^2-m_W^2} \, {\overline\Psi}\gamma^\mu (1-\gamma_5) \Psi~,
\eeq
where $q^2$ is the momentum transfer (energy scale) of the interaction.
We can expand the $W$-propagator in powers of $q^2/m_W^2$:
\beq
{1\over q^2-m_W^2} = -{1\over m_W^2} \left[ 1 + {q^2\over m_W^2} +
...\right].
\eeq
The interaction of Eq.~(\ref{SMCC}) can thus be written as the sum of an
infinite number of terms. However, we note that, for energies well below
the $W$ mass, only the first term is important. This is simply the 4-fermion
interaction of the Fermi theory:
\beq
-{G_F\over\sqrt{2}}\, {\overline\Psi}\gamma_\mu (1-\gamma_5) \Psi \,
{\overline\Psi}\gamma^\mu (1-\gamma_5) \Psi~,
\eeq
where $G_F/\sqrt{2} = g^2/8m_W^2$. In other words, the Fermi theory is the
effective theory produced when one ``integrates out'' the heavy degrees of
freedom (in this case, the $W$ boson). It is valid at energy scales much
less than the scale of heavy physics ($q^2 \ll m_W^2$).

Note that, as $q^2$ approaches $m_W^2$, one can no longer truncate after the
lowest-order term in $q^2/m_W^2$. This is evidence that the effective
Lagrangian is breaking down -- each of the infinite number of terms becomes
equally important as one is approaching energy scales where
the heavy degrees of freedom can be directly produced, {\elevenit i.e.}
they cannot be
integrated out. Note also that the truncated effective Lagrangian (the
Fermi theory) violates $S$-matrix unitarity for $q^2>m_W^2/(g^2/4\pi)$.
Unitarity is
restored in the full theory by propagator (form factor) effects and the
scale at which unitarity is apparently violated gives an upper bound for
the masses of the heavy degrees of freedom (here the $W$ mass). In a
weakly coupled theory like the SM this upper bound substantially
overestimates the masses of the heavy degrees of freedom.
Apart from resonance enhancement one needs strong interaction dynamics
to obtain cross sections in the full theory which approach the unitarity
limits. As the energy scale is increased, new channels will open up in
addition ({\elevenit e.g.} $WW$ and $WZ$ production in the case of the
Fermi theory).
However, the cross sections of these new channels may be too low to be
observable, especially if the underlying dynamics is perturbative in nature.
These features, which are easily understood in the context of the SM and the
Fermi theory, are general properties of all effective Lagrangians.

\vglue 0.2cm
{\elevenit\noindent 2.2 Power Counting}
\vglue 0.2cm

In order to define an effective Lagrangian, it is necessary to specify the
symmetry and the particle content of the low-energy theory. Since  all
experimental evidence is consistent with the existence of an  $SU(2)_L
\times U(1)_Y$ gauge symmetry it is natural to require the effective
Lagrangian
describing anomalous TGV's to possess this invariance. Inspecting
Eq.~(\ref{LeffWWV}), the phenomenological effective Lagrangian  ${\cal
L}_{eff}^{WWV}$ describing anomalous $WWV$ couplings appears not to
respect this constraint. This  impression is
wrong, however, since Eq.~(\ref{LeffWWV}) can be interpreted  as the
unitary-gauge expression of an effective Lagrangian in which the  $SU(2)_L
\times U(1)_Y$ gauge symmetry is manifest~\cite{BurLon}. How this symmetry
is  realized depends on the particle content of the effective Lagrangian.
If one  includes a Higgs boson, the symmetry can be realized linearly,
otherwise a  nonlinear realization of the gauge symmetry is required. We
will discuss each  of these options in turn.

\vglue 0.2cm
\noindent
$\bullet$ Linear Realization
\vglue 0.2cm

We first consider the linear realization scenario, in which a Higgs doublet
field $\Phi$ is included in the low-energy particle content. This is also
called the ``decoupling physics'' scenario in the literature because,
with the
inclusion of a light Higgs boson, the scale of new physics is allowed to be
arbitrarily large, even $\Lambda \sim 10^{15}$~GeV would be
self-consistent.
In addition to the Higgs field the building blocks of the effective
Lagrangian are covariant derivatives of the Higgs field, $D_\mu \Phi$,
and the field strength tensors $W_{\mu\nu}$ and $B_{\mu\nu}$ of the $W$
($SU(2)_L$) and the $B$ ($U(1)_Y$) gauge fields:
\begin{eqnarray}\label{Wmunu}
[D_\mu,D_\nu] = \hat{B}_{\mu\nu} + \hat{W}_{\mu\nu} =
                i\,{g'\over 2}\,B_{\mu\nu} +
                i\,g\,{\sigma^a\over 2}\,W^a_{\mu\nu}\; .
\end{eqnarray}
Here, $\sigma^a$, $a=1,\dots,3$ denote the Pauli matrices, and $g$ and
$g'$ are the $SU(2)_L$ and $U(1)_Y$ gauge coupling constants,
respectively.
Considering dimension-6 operators only, 11~independent such operators can
be constructed~\cite{bw,HISZ} of which only~7 are relevant for our
discussion:
\begin{eqnarray}\label{LRoperators}
{\cal L}_{eff} & = & \sum_{i=1}^7 {f_i\over \Lambda^2}\, {\cal O}_i =
{1\over \Lambda^2} \biggl( 
f_{\Phi,1}\; (D_\mu\Phi)^\dagger \Phi\; \Phi^\dagger (D^\mu \Phi)\;
+\; f_{BW}\; \Phi^\dagger\hat{B}_{\mu\nu}\hat{W}^{\mu\nu} \Phi\;
\nonumber \\[-1.mm]
& + & f_{DW}\; Tr([D_\mu, \hat{W}_{\nu\rho}]\, [D^\mu,\hat{W}^{\nu\rho}])\;
-\; f_{DB}\; {g'^2\over 2}\,(\partial_\mu B_{\nu\rho})(\partial^\mu
B^{\nu\rho})\; \nonumber  \\
& + & f_B\; (D_\mu\Phi)^\dagger\hat{B}^{\mu\nu}(D_\nu\Phi)\;
+\; f_W\; (D_\mu\Phi)^\dagger\hat{W}^{\mu\nu}(D_\nu\Phi)\; \nonumber \\
& + & f_{WWW}\; Tr[\hat{W}_{\mu\nu}\hat{W}^{\nu\rho}{\hat{W}_\rho}\,^\mu]
\biggr) \; .
\end{eqnarray}

The first four operators, ${\cal O}_{\Phi,1}$, ${\cal O}_{BW}$,
${\cal O}_{DW}$, and ${\cal O}_{DB}$,
affect the gauge boson two-point functions at tree level~\cite{gw} and as a
result the coefficients of these four operators are
severely constrained by present low energy data.
The remaining three, ${\cal O}_B$, ${\cal O}_W$ and ${\cal O}_{WWW}$, give
rise to non-standard triple gauge boson couplings. Their presence in the
effective Lagrangian leads to deviations of the $WWV$ couplings from the SM,
namely~\cite{HISZ,deR}
%
\begin{eqnarray}\label{kapanom}
\Delta\kappa_\gamma = (f_B+f_W)\;{m_W^2\over 2 \Lambda^2}\; , \;&&\;\;
\Delta\kappa_Z = \left(f_W-s^2(f_B+f_W)\right)\;
                 {m_Z^2\over 2 \Lambda^2}\; , \\
\Delta g_1^Z = f_W\;{m_Z^2\over 2 \Lambda^2} & = &
\Delta\kappa_Z + {s^2\over c^2}\,\Delta\kappa_\gamma \; , \label{kapanomc}\\
\lambda_\gamma = \lambda_Z = \lambda & = &
{3m_W^2g^2\over 2 \Lambda^2}\; f_{WWW}\;
 ,\label{kapanomd}
\end{eqnarray}
%
with $s = \sin\theta_W$ and $c = \cos\theta_W$ . Note that all
anomalous TGV's are suppressed
by a factor $m_W^2/\Lambda^2$ and hence they vanish in the decoupling
limit.  In fact this behaviour is required by unitarity considerations
with,  typically, $|f_i| \lsim 32\pi$~\cite{bz}. In general, the
coefficients  $f_i$ are expected to be numbers of order unity. Hence,
taking  $\Lambda \sim 1$~TeV, one might expect anomalous TGV's of ${\cal
O}(10^{-2})$.
As  pointed out by Einhorn and collaborators~\cite{arzt}, the dimension six
operators ${\cal O}_{WWW}$, ${\cal O}_W$ and ${\cal O}_B$ which lead to
anomalous TGV's cannot be generated at the tree level by any
renormalizable underlying theory which leads to the effective Lagrangian
of Eq.~(\ref{LRoperators}). Thus, in this scenario, the expected
size of the anomalous TGV's would be tiny, $\sim 1/(16\pi)^2\;
(m_W^2/\Lambda^2)$, and only small scales $\Lambda$ would be accessible
experimentally. In the same scenario dimension~8 operators leading to TGV
contributions can be generated at tree level and, thus, they might dominate
over the dimension~6 terms considered above if $\Lambda$ is sufficiently
small. Since the correlations between different anomalous $WWV$ couplings
exhibited in  Eqs.~(\ref{kapanomc}) and~(\ref{kapanomd}) are due to the
truncation
of the effective Lagrangian at the dimension six level~\cite{HISZ} these
relationships would not even be approximately correct in this case.

Anomalous $Z\gamma V$ couplings originate only from terms of dimension~8
or higher in the effective Lagrangian and, therefore, are expected to be
${\cal O}(m_Z^4/\Lambda^4)$.

\vglue 0.2cm
\noindent
$\bullet$ Nonlinear Realization
\vglue 0.2cm

Let us now turn to the scenario in which the $SU(2)_L \times U(1)_Y$ gauge
symmetry is realized non-linearly (``non-decoupling physics"). In this case,
one includes only the would-be Goldstone bosons (WBGB's) which give
masses to the $W$- and $Z$-bosons. Since there is no Higgs boson, the
low-energy Lagrangian violates unitarity at a scale of roughly $4\pi v
\sim 3$ TeV, so that the new physics must appear at a scale
$\Lambda \lsim 4 \pi v$.

A number of nonlinear realizations appear in the literature, all of
which are similar~\cite{nonlin}. For the purpose of illustration we will
choose one which conserves the
custodial $SU(2)_C$ symmetry of the SM in the limit $g'\to 0$.
Using the matrix $\Sigma \equiv \exp(i {\vec\omega}\cdot
{\vec\sigma}/v)$, where the $\omega_i$ are the WBGB's, we define
the $SU(2)_L\times U(1)_Y$ covariant derivative:
\beq
D_\mu \Sigma \equiv \partial_\mu \Sigma + {i\over 2}\,g W^a_\mu\sigma^a
\Sigma - {i\over 2}\, g' B_\mu \Sigma \sigma_3~.
\eeq
One then constructs terms in the effective Lagrangian using field strengths
($W_{\mu\nu}$, $B_{\mu\nu}$) and covariant derivatives. This effective
Lagrangian is known generically as a ``chiral Lagrangian,'' due to its
similarity to low-energy QCD (and chiral perturbation theory).
In the unitary gauge the covariant derivative becomes a linear
combination of gauge bosons. Thus, a gauge-boson field can be constructed
by taking the trace of $D_\mu \Sigma$ with the appropriate $\sigma$ matrix,
{\elevenit e.g.} $Z_\mu \sim {\elevenrm Tr}[\sigma_3 \Sigma^\dagger
D_\mu \Sigma]$. In
this way, we can write the Lagrangian of Eq.~(\ref{LeffWWV}) in terms of
$SU(2)_L\times U(1)_Y$-invariant quantities.

Our experience with QCD tells us how to estimate the size of any term in a
chiral Lagrangian. This estimate is called ``naive dimensional analysis''
(NDA)~\cite{NDA}. It states that a term having $b$ WBGB fields, $f$
(weakly-interacting) fermion fields, $d$ derivatives and $w$ gauge fields
has a coefficient whose size is
\beq  c_n(\Lambda) \sim v^2 \Lambda^2 \; \left( { 1\over v} \right)^b \;
\left(
{ 1\over \Lambda^{3/2}} \right)^f \; \left( { 1\over \Lambda} \right)^d \;
\left( { g\over \Lambda} \right)^w.
\eeq
Applying NDA to the terms in Eqs.~(\ref{LeffWWV}) and~(\ref{anomTGVs}),
we see that $\Delta g_1^V$ and  $\Delta\kappa_V$ are of ${\cal
O}(m_W^2/\Lambda^2)$. In
other words, just as in the linear realization, these terms are effectively
of dimension 6 (in the sense that there is an explicit factor of
$1/\Lambda^2$).  On the other hand, we see that the
$W^{\dagger}_{\rho\mu}{W^{\mu}}_{\nu}V^{\nu\rho}$ term is effectively of
dimension~8, {\elevenit i.e.} the coefficient $\lambda_V$ is expected
to be of order
$m_W^4/\Lambda^4$. Thus, within the nonlinear realization scenario, the
$\lambda_V$ terms are expected to be negligible compared to those
proportional to $\Delta g_1^V$ and $\Delta\kappa_V$.

Within the nonlinear realization scenario, there are two operators which
contribute to anomalous TGV's (and not to two-point functions) at lowest
order \cite{chiralops,DawVal}. Writing the heavy mass dependence
explicitly, they are:
\beq
\label{NLRoperators}
-i g\, {v^2\over \Lambda^2}\, L_{9L}\, {\elevenrm Tr}\left[ W^{\mu\nu}
D_\mu \Sigma D_\nu \Sigma^\dagger \right]
-i g'\, {v^2\over \Lambda^2}\, L_{9R}\, {\elevenrm Tr}\left[ B^{\mu\nu}
D_\mu \Sigma^\dagger D_\nu \Sigma \right].
\eeq
These are related to $\Delta g_1^V$ and $\Delta\kappa_V$ by:
\begin{eqnarray}
\Delta g_1^Z & = & {e^2\over 2c^2 s^2}\, {v^2\over \Lambda^2}\, L_{9L} ,
\nonumber \\
\Delta \kappa_\gamma & = & {e^2\over 2 s^2}\, {v^2\over \Lambda^2}\,
(L_{9L} + L_{9R} ), \\
\Delta\kappa_Z & = & {e^2\over 2c^2 s^2}\, {v^2\over \Lambda^2}\, (L_{9L}
c^2 - L_{9R} s^2), \nonumber
\end{eqnarray}
where $s^2 c^2 = \pi \alpha(m_Z) / \sqrt{2} G_F m_Z^2$. (The $g_5^Z$
coupling is studied in Ref.~\cite{g5}.) Note that as  far as these three
TGV's are concerned the linear and the nonlinear realization are obtained
from each other by identifying $L_{9L}=2f_W$ and $L_{9R}=2f_B$. In
particular the correlation between TGV's as given in
Eqs.~(\ref{kapanomc}) and~(\ref{kapanomd}) holds in both frameworks as
long as higher-dimensional operators can be neglected.

Anomalous $Z\gamma V$ couplings again originate only from higher order
terms in the effective Lagrangian.

\vglue 0.2cm
{\elevenit\noindent 2.3 Form Factors}
\vglue 0.2cm

Although the anomalous TGV's $\Delta g_1^V$, $\Delta\kappa_V$,
{\elevenit etc.} appear
as constants in Eqs.~(\ref{anomTGVs}) and~(\ref{EQ:ZgammaV}), they should
rather be considered as form factors. Consider the $\Delta\kappa_V$ term,
$W^{\dagger}_{\mu}W_{\nu}V^{\mu\nu}$. One can write down similar
higher-order terms such as
\beq
{1\over \Lambda^2}\, W^{\dagger}_{\mu}W_{\nu} \, \Square V^{\mu\nu}~,
\eeq
which has the same Feynman rules as
$W^{\dagger}_{\mu}W_{\nu}V^{\mu\nu}$, except for a multiplicative $q^2$
dependence due to the derivatives in $\Square$. Taking into account all
such operators, the overall coefficient of the Feynman rule is not
$\Delta\kappa_V$, but rather a form factor
\beq
\Delta\kappa_V^0 \left( 1 + a {q^2\over \Lambda^2} +
b \left({q^2\over \Lambda^2}\right)^2 + ... \right)\; ,
\eeq
where $a$, $b$, {\elevenit etc.} are ${\cal O}(1)$. Since a constant
anomalous TGV would lead
to unitarity violation at high energies~\cite{joglekar} such a form factor
behaviour is a feature of any model of anomalous couplings. When studying
$W^+W^-$ production at an $e^+e^-$ collider at fixed $q^2=s$ this form
factor behaviour is of no consequence.
Weak boson pair production at hadron colliders, however, probes the TGV's
over a large $q^2=\hat s$ range and is very sensitive to the fall-off of
anomalous TGV's which necessarily happens once the threshold of new physics
is crossed.
Not taking this cutoff into account results in unphysically large cross
sections at high energy (which violate unitarity) and thus leads to a
substantial overestimate of experimental sensitivities. In our analysis
in Section~3 we will assume a simple power law behaviour, {\elevenit
e.g.}
\begin{equation}\label{FF}
\Delta\kappa_V(q^2) = { \Delta\kappa_V^0 \over (1+q^2/\Lambda_{FF}^2)^n }\; ,
\end{equation}
and similarly for the other TGV's. Here $\Lambda_{FF}$ is the form
factor scale which is a function of the scale of new physics, $\Lambda$.
For $WWV$ couplings we shall use the exponent $n=2$,
which will be referred to as the `dipole form factor' below. For
$Z\gamma V$ couplings we choose $n=3$ ($n=4$) for $h_3^V$ ($h_4^V$). Due
to the form factor behaviour of the anomalous couplings, the
experimental limits extracted from hadron collider experiments
explicitly depend on $\Lambda_{FF}$.

The values $\Delta\kappa_V^0$ {\elevenit etc.} of the form factors at
low energy are constrained by partial wave unitarity of the inelastic
vector boson pair production amplitude in fermion antifermion
annihilation at arbitrary center of mass energies. Assuming that only
one anomalous coupling is nonzero at a time, one finds, for
$\Lambda_{FF}\gg m_W,\,m_Z$~\cite{bz,UBELB2}
\begin{eqnarray}
|\Delta g_1^{Z0}| \leq  {n^n\over (n-1)^{n-1}}\, {0.84~{\rm
TeV}^2\over \Lambda_{FF}^2}\,,~~~~~~~~ & &
|g_5^{Z0}| \leq  {(2n)^n\over (2n-1)^{n-1/2}}\, {3.2~{\rm
TeV}\over \Lambda_{FF}}\,,\label{EQ:unitbf} \\[1.mm]
|\Delta\kappa_\gamma^0| \leq  {n^n\over (n-1)^{n-1}}\, {1.81~{\rm
TeV}^2\over \Lambda_{FF}^2}\,,~~~~~~~~~ & &
|\Delta\kappa_Z^0|  \leq  {n^n\over (n-1)^{n-1}}\, {0.83~{\rm
TeV}^2\over \Lambda_{FF}^2}\,, \\[1.mm]
|\lambda_\gamma^0| \leq  {n^n\over (n-1)^{n-1}}\, {0.96~{\rm
TeV}^2\over \Lambda_{FF}^2}\,,~~~~~~~~~~~ & &
|\lambda_Z^0| \leq  {n^n\over (n-1)^{n-1}}\, {0.52~{\rm
TeV}^2\over \Lambda_{FF}^2}\,, \\[1.mm]
\left\vert h_{30}^Z\right\vert  \leq
{\left({\scriptstyle{2\over 3}}\, n\right)
^n\over\left({\scriptstyle{2\over 3}}\, n-1\right)^{n-3/2}}~{0.126~{\rm
TeV}^3\over\Lambda_{FF}^3}\,,~~~~~\! & &
\left\vert h_{30}^\gamma\right\vert  \leq
{\left({\scriptstyle{2\over 3}}\, n\right)
^n\over\left({\scriptstyle{2\over 3}}\, n-1\right)^{n-3/2}}~{0.151~{\rm
TeV}^3\over\Lambda_{FF}^3}\,, \\[1.mm]
\left\vert h_{40}^Z\right\vert  \leq  {\left({\scriptstyle{2\over
5}}\, n\right)
^n\over\left({\scriptstyle{2\over 5}}\, n-1\right)^{n-5/2}}~{2.1\cdot
10^{-3}~{\rm TeV}^5\over\Lambda_{FF}^5}\,, & &
\left\vert h_{40}^\gamma\right\vert  \leq
{\left({\scriptstyle{2\over 5}}\, n\right)
^n\over\left({\scriptstyle{2\over 5}}\, n-1\right)^{n-5/2}}~{2.5\cdot
10^{-3}~{\rm TeV}^5\over\Lambda_{FF}^5}\,.~~
\label{EQ:unitbl}
\end{eqnarray}
The bounds listed in Eqs.~(\ref{EQ:unitbf}) --~(\ref{EQ:unitbl}) have
been computed with $m_W=80$~GeV and $m_Z=91.1$~GeV. In order to satisfy
unitarity, $n\geq 1$ for $\Delta g_1^{Z}$, $\Delta\kappa_V$ and
$\lambda_V$, $n\geq 1/2$ for $g_5^Z$, $n\geq 3/2$ for $h_3^V$, and
$n\geq5/2$ for $h_4^V$. If more than one coupling is varied at a time,
cancellations between the TGV's may occur, and the unitarity limits are
weakened somewhat. For $\Lambda_{FF}\gg m_W,\, m_Z$, the unitarity
limits drop like a power of $1/\Lambda_{FF}$ with increasing values of
the form factor scale. The experimental limits
obtained from hadron collider experiments must be compared with the
bounds derived from $S$-matrix unitarity. Experiments constrain the
$WWV$ and $Z\gamma V$ couplings non-trivially only if the experimental
limits are more stringent than the unitarity bounds, for a given value of
$\Lambda_{FF}$.

Strictly speaking the appearance of form factor effects implies that the
effective Lagrangian description in terms of a small set of low-dimensional
operators breaks down, {\elevenit i.e.} one is probing weak boson pair
production at
the scale of new physics. New channels are expected to open up as well.
However, the corresponding cross sections might be too small to be
observable immediately or the experimental signatures might be obscured
by backgrounds  (compare {\elevenit e.g.} $WZ$ production in the Fermi
theory). Thus
form factors are a tool to extend the use of effective Lagrangians to
the entire energy range which is accessible at hadron colliders.

\vglue 0.2cm
{\elevenit\noindent 2.4 Phenomenological Bounds from High Precision
Experiments}
\vglue 0.2cm

In Section~2.2, we have discussed the reasons why anomalous TGV's are
expected to be ${\cal O}(m_W^2/\Lambda^2)$ at most in an effective
Lagrangian approach. However, it is also interesting to ask
what is known about anomalous TGV's from experiment. The errors of present
direct measurements, via pair production of electroweak bosons, are still
very large (of order 100\%, see Section 3). More precise constraints might
then arise from loop contributions to precisely measured quantities such as
$(g-2)_\mu$~\cite{arzt,muf}, the $b\to s\gamma$ decay
rate~\cite{btosgamma,CLEO}, $B\to K^{(*)}\mu^+\mu^-$~\cite{Baillie}, the
$Z\to b\bar b$~\cite{Eboli} rate
and oblique corrections ({\elevenit i.e.} corrections to the two point
functions) to 4-fermion $S$-matrix elements. Oblique corrections
combine information from the recent LEP/SLD data, neutrino scattering
experiments, atomic parity violation, $\mu$-decay, and the $W$-mass
measurement at hadron colliders.
These analyses have been performed for $WWV$ couplings in the context of
linear and nonlinear realizations, and we discuss both of these in turn.

\vglue 0.2cm
\noindent
$\bullet$ Linear Realization
\vglue 0.2cm

A complete analysis of low energy constraints on the coefficients of the
effective Lagrangian of Eq.~(\ref{LRoperators}) was performed in
1992~\cite{HISZ}. Here we update these results by using the comprehensive
1994 analysis of electroweak data by Hagiwara {\elevenit et al.}~\cite{HMHK}.
With $\alpha = 4\pi \overline{e}^2(0)$ taken as an input parameter,
the neutral- and charged-current data may be parameterized in terms of three
effective form-factors, $\bar g_Z^2(q^2)$ and $\bar g_W^2(q^2)$ defining
the coupling strength of the $Z$ and the $W$ at momentum transfer $q$ and
the square of the effective weak mixing angle, $\bar s^2(q^2)$. For
$m_t = 174$~GeV and $\alpha_s(m^2_Z)=0.12$ the LEP and SLD data can be
summarized in terms of
%
%
\begin{eqnarray}\label{gzmzdata}
\bar g_Z^2(m_Z^2)  =  0.55673 \pm 0.00087\; ,  \qquad
\bar s^2(m_Z^2)  =  0.23051 \pm 0.00042  \; , \qquad
\rho = 0.28\; ,
\end{eqnarray}
where $\rho$ is the correlation of the two values.
In a similar fashion the low-energy data on neutrino scattering and atomic
parity violation determine the same form-factors at zero momentum transfer:
%
%
\begin{eqnarray} \label{gz0data}
\bar g_Z^2(0)  =  0.5462 \pm 0.0036  \;, \qquad
\bar s^2(0)  =  0.2353 \pm 0.0044  \;, \qquad
\rho=0.53 \; .
\end{eqnarray}
Finally, the $W$-mass measurement at hadron colliders together with the input
value of $G_F$ can be translated into a measurement of $\bar g_W^2(0)$:
%
%
\begin{eqnarray}
\bar g_W^2(0) = 0.4225 \pm 0.0017  \;. \label{gw0data}
\end{eqnarray}

These five measurements are closely related to other formulations of the
oblique corrections, like the $S$, $T$, and $U$ parameters of Peskin and
Takeuchi~\cite{STU}. The new feature here is the inclusion of the $q^2$
dependence of the
form -- factors~\cite{HISZ,HMHK,maksymyk}. Indeed, new physics contributions
like the operators ${\cal O}_{DW}$ or ${\cal O}_{DB}$ do lead to a
nontrivial $q^2$ dependence of the form-factors, and the more general
analysis is needed to constrain these operators. Low energy bounds are
obtained by fitting
\begin{eqnarray}
S &=& S_{SM}(m_t,m_H) + \Delta S\; , \\
T &=& T_{SM}(m_t,m_H) + \Delta T\;\;\; etc.
\end{eqnarray}
to the data. Here the SM contributions ($S_{SM}$ {\elevenit etc.})
introduce a significant
dependence on the values of the Higgs boson and the top quark masses.

The four operators ${\cal O}_{DW}$, ${\cal O}_{DB}$, ${\cal O}_{BW}$, and
${\cal O}_{\Phi,1}$ contribute already at tree level,
\begin{eqnarray}\label{resff}
\Delta\delta\rho & = & \alpha \Delta T =
- {v^2\over 2\Lambda^2}\; f_{\Phi,1}\; , \\
\Delta S & = & - 32\pi s^2\,{m_W^2\over \Lambda^2}\; (f_{DW}+f_{DB})
               -4\pi\,{v^2\over\Lambda^2}f_{BW}\; ,\label{resffb}
\end{eqnarray}
with similar results for the other form-factors. Fitting these to the five
data points one obtains measurements of the coefficients of the operators in
the effective Lagrangian. The central values depend on the top quark
and Higgs boson masses which we parameterize in terms of
\begin{equation}\label{xtxh}
x_t = {m_t-175\; {\rm GeV}\over 100\; {\rm GeV} }\; ,\qquad
x_H = {m_H \over 100\; {\rm GeV} }\; .
\end{equation}
Within better than 5\% of the $1\sigma$ errors, and in the ranges
140~GeV~$<m_t  < $~220~GeV and 60~GeV~$ < m_H < $~800~GeV  this dependence
is given by
\begin{eqnarray}\label{valfi}
f_{DW} & = & -0.35 + 0.012\, {\rm log}\, x_H - 0.14\, x_t \pm 0.62  \; ,\\
f_{DB} & = & -11 + 0.11\, {\rm log}\, x_H - 0.58\, x_t \pm 11 \; ,\\
f_{BW} & = & 3.1  + 0.072\,{\rm log}\,x_H + 0.22\, x_t \pm 2.6  \; ,\\
f_{\Phi,1} & = & 0.23 -0.031\, {\rm log}\, x_H + 0.36\, x_t \pm 0.17 \; ,
\end{eqnarray}
assuming $\Lambda=1$~TeV. The correlation matrix $C$ is found to be
\begin{eqnarray}\label{corrmatr}
C = { V_{ij} \over \sqrt{ V_{ii} V_{jj} } } &=& \left(
\begin{array}{c@{\qquad}c@{\qquad}c@{\qquad}c}

 1.   &  $-$0.323 &  0.151    &  $-$0.228 \;\; \\
          &  1.       &  $-$0.979 &  $-$0.806 \;\; \\
          &               &  1.       &   0.905 \;\; \\
          &               &               &   1.

\end{array} \right) \: .
\end{eqnarray}
Both the $1\sigma$ errors and the correlation matrix elements are
independent of $m_H$ and $m_t$ to high precision. Note the strong
correlations between the coefficients of the dimension six operators,
in particular between $f_{DB}$, $f_{BW}$ and $f_{\Phi,1}$.
While the contributions of these four operators are already constrained
at the tree level, the anomalous $WWV$ couplings only contribute at the
1-loop level
to the oblique correction parameters. Neglecting all terms which are not
logarithmically divergent for $\Lambda\to\infty$, the leading effects are
given by replacing $f_{DW}$ {\elevenit etc.} in Eqs.~(\ref{resff})
and~(\ref{resffb}) by the renormalized quantities
%
\begin{eqnarray}\label{firen}
f_{DW}^r & = & f_{DW} -
        {1\over 192\pi^2}\,f_W\; {\rm log}{\Lambda^2\over \mu^2}\; ,
        \\[1.mm]
f_{DB}^r & = & f_{DB} -
        {1\over 192\pi^2}\,f_B\; {\rm log}{\Lambda^2\over \mu^2}\; ,
        \\[1.mm]
f_{BW}^r & = & f_{BW} + {\alpha\over 32\pi s^2}\,{\rm
log}{\Lambda^2\over\mu^2}
\Biggl(f_B\left({20\over3}+{7\over 3c^2}+{m_H^2\over m_W^2}\right)
\nonumber  \\[1.mm]      &&
\qquad \qquad -\; f_W\left (4+{1\over c^2}-{m_H^2\over m_W^2}\right )
\; + \; 12g^2f_{WWW}  
\Biggr)\; ,  \\[1.mm]
f_{\Phi,1}^r & = & f_{\Phi,1} + {3\alpha\over 8\pi c^2}\,
{\rm log}{\Lambda^2\over\mu^2}
\left(f_B{m_H^2\over v^2}+{3m_W^2\over v^2}\,(f_B+f_W)\right)\; .
\label{firend}
\end{eqnarray}
%
Here, $\mu$ denotes the unit-of-mass of the dimensional regularization
which has been used to regulate the divergencies which appear in the
calculation.
The ${\rm log}{\Lambda^2\over\mu^2}$ terms in Eqs.~(\ref{firen})
--~(\ref{firend})
describe mixing of the operators between the new physics scale $\Lambda$ and
the weak boson mass scale $\mu=m_W$.

The results of Eqs.~(\ref{firen}) --~(\ref{firend}) nicely illustrate
the problem
of deriving low energy bounds for the TGV's. The dominant contributions of
the coefficients $f_B$, $f_W$ and $f_{WWW}$ are merely renormalizations of
those 4 operators which already contribute at tree level. Also, the
precision electroweak data are barely sufficient to constrain all four
coefficients $f_{DW}^r$, $f_{DB}^r$, $f_{BW}^r$, and $f_{\Phi,1}^r$.
Hence, indirect bounds
on the TGV's are only possible if one makes stringent assumptions on
the size
of these `tree level' coefficients. An analogous problem appears when
considering 1-loop contributions of the TGV's to $(g-2)_\mu$, $Z\to
b\bar b$ or $b\to s\gamma$ and hence data on those observables do not
provide model-independent bounds either.

\begin{figure}[t]
\vskip 9.cm
\includegraphics{fig1.ps}
{\sl\noindent Figure~1: Constraints on a) $\Delta\kappa_\gamma$ vs.
$\lambda$ and b)
$\Delta\kappa_Z$ vs. $\lambda$ at 95\% confidence level (CL). All
coefficients of the dimension
six operators in Eq.~(\ref{LRoperators}) are assumed to vanish except for
a) $f_B = f_W$ and $f_{WWW}$ and b) $f_B = -f_W$ and $f_{WWW}$.
Correlations are shown for three representative Higgs boson masses. }
\end{figure}
Nevertheless, one may proceed and $assume$ that cancellations between tree
level and 1-loop contributions or between any of the coefficients of the
dimension~6 operators are unnatural. In practice one assumes that all $f_i$
vanish at the scale of new physics $\Lambda$ except for the one or two whose
effect one wants to analyse. The result of such an exercise is shown in
Fig.~1.
Clearly anomalous TGV's of ${\cal O}(1)$ are still allowed by the data.
Note that these bounds become more stringent as the Higgs boson mass
increases,
pointing to more severe bounds in the nonlinear realization scenario. If
the top quark mass is varied between 150~GeV and 200~GeV, the range
allowed for the anomalous couplings increases by up to 50\%.

Other processes which at the 1-loop level are sensitive to anomalous
gauge boson couplings also give constraints of ${\cal O}(1)$ at best.
The current CLEO measurement of the inclusive $b\to s\gamma$ decay
rate~\cite{CLEO}, for example, still allows $-2.6<\Delta\kappa_\gamma
<-1.2$ and $-0.5<\Delta\kappa_\gamma<0.4$ (for $\lambda_\gamma=0$),
and $-1.7<\lambda_\gamma<1.0$ (for $\Delta\kappa_\gamma=0$) at 95\% CL.

A more stringent assumption on the coefficients of the dimension~6
operators has been proposed by De~R\'ujula {\elevenit
et al.}~\cite{deR}. There are no obvious symmetries which distinguish
the tree level
operators ${\cal O}_{BW}$, ${\cal O}_{DW}$, ${\cal O}_{DB}$, and ${\cal
O}_{\Phi,1}$ from the remaining
ones. For a generic model of the underlying dynamics one may hence expect
{\elevenit e.g.} $|f_B+f_W| \approx |f_{BW}|$ which with
$f_{BW}/\Lambda^2 = (3.1
\pm 2.6)~{\rm TeV}^{-2}$ implies $|\Delta\kappa_\gamma|= |f_B+f_W|\;
{m_W^2 /2\Lambda^2} < 0.03$ at ``95\% CL'', a value too small to be
observable in $W^+W^-$ production at LEP~II, but still in the
interesting range for future
linear colliders. Although this naturalness argument is compelling,
it is clearly not a proof that anomalous TGV's are indeed small.

\vglue 0.2cm
\noindent
$\bullet$ Nonlinear Realization
\vglue 0.2cm

There are several analyses in the literature which discuss the bounds on
anomalous TGV's in the context of the nonlinear realization
scenario~\cite{DawVal,BGKLM,otherTGV}. All conclusions are quite
similar.
The limits obtained in the nonlinear realization framework are very similar
to those obtained in the linear realization scenario for a large Higgs
boson mass. In the following, we will briefly review the results of
Ref.~\cite{BGKLM}.

As in the linear realization case the effective Lagrangian is
nonrenormalizable, and therefore the loop diagrams diverge.
Conceptually, this is not a
problem -- the effective Lagrangian already contains an infinite number of
terms, so one can just add a counterterm to cancel the infinities found in
any loop calculation. In other words, if an anomalous TGV contributes at
loop level to an observable, the divergence of the calculation just
renormalizes the coefficient of that observable. At the calculational
level, however, one has to decide how to deal with the infinities. In the
past it was common to simply use a cutoff $\tilde\Lambda$ to regulate the
divergence. With this technique one often obtained a cutoff dependence of
the form $\tilde\Lambda^2$ or even $\tilde\Lambda^4$, resulting in extremely
stringent constraints on anomalous TGV's. However, it was argued in
Ref.~\cite{BurLon} that the use of cutoffs was incorrect, and often gave
misleading results. Instead the authors of Ref.~\cite{BurLon} advocate the
use of dimensional regularization, along with the decoupling-subtraction
renormalization scheme. This is the procedure used in Ref.~\cite{BGKLM}.

For the calculational details, we refer the reader to Ref.~\cite{BGKLM}.
Here we only present the results of the global fit. First, consider the
case in which
only one of the anomalous TGV couplings, $\Delta g_1^Z$, $\Delta\kappa_V$
and $\lambda_V$, is nonzero. (The coupling $g^Z_5$ was not considered in
this paper.) The fit gave the following constraints at $1\sigma$:
\begin{eqnarray}
\label{results}
\Delta g_1^Z & = & -0.033 \pm 0.031, \nonumber \\
\Delta\kappa_\gamma & = & 0.056 \pm 0.056, \nonumber \\
\Delta\kappa_Z & = & -0.0019 \pm 0.044, \\
\lambda_\gamma & = & -0.036 \pm 0.034, \nonumber \\
\lambda_Z & = & 0.049 \pm 0.045. \nonumber
\end{eqnarray}
If taken at face value, these limits would imply that most anomalous
TGV's are
too small to be seen at LEP~II or in future Tevatron experiments (see
Section~3). The LHC and NLC, on the other hand, will be able to improve
these bounds considerably.

However, one should keep in mind that these bounds are rather artificial.
It is very hard to
imagine that physics beyond the SM will produce only one anomalous TGV. In
general, all such couplings will be produced. In a fit to all five
anomalous couplings simultaneously, the constraints virtually
disappear, due to
the possibility of cancellations. At best, one can only conclude that the
anomalous TGV couplings are less than ${\cal O}(1)$ and even here one must
assume that tree level contributions do not cancel the TGV effects.

Even so, the bounds of Eq.~(\ref{results}) are interesting. These values
represent the sensitivity of the global fit of electroweak data to specific
anomalous couplings. Once all of the couplings are allowed to vary
simultaneously, no significant bound remains. This obviously implies that,
in that part of the allowed region for which the TGV couplings are large,
cancellations occur among the contribution of the various anomalous
couplings to low-energy observables. Equation~(\ref{results}) gives an
indication
of the level of cancellation required to account for the low-energy data in
the event that an anomalous TGV at the 10\% level is discovered.

\vglue 0.2cm
{\elevenit\noindent 2.5 Summary}
\vglue 0.2cm

We have discussed our theoretical understanding of, and the
phenomenological constraints from high precision experiments on, the
anomalous TGV's of Eqs.~(\ref{LeffWWV})
and~(\ref{anomTGVs}). The phenomenological effective Lagrangian
describing anomalous couplings appears not to respect the $SU(2)_L \times
U(1)_Y$ gauge symmetry. However, it can be interpreted as the
unitary-gauge expression of an effective Lagrangian in which the
$SU(2)_L \times
U(1)_Y$ symmetry is manifest. How this symmetry is realized depends on
the particle content of the effective Lagrangian.
Regardless of how one realizes the $SU(2)_L \times
U(1)_Y$ gauge symmetry, the anomalous TGV's $\Delta g_1^Z$,
$\Delta\kappa_V$, $\lambda_V$ and $g_5^Z$ are expected to be at most
${\cal O}(m_W^2/\Lambda^2)$, where $\Lambda$ is the scale of new
physics. (In the nonlinear-realization scenario, $\lambda_V$ is ${\cal
O}(m_W^4/\Lambda^4)$). $Z\gamma V$ couplings are at most ${\cal O}(m_Z^4
/\Lambda^4)$. Thus, for $\Lambda\sim 1$~TeV anomalous TGV's of
${\cal O}(10^{-2})$ or less are expected. The
discovery of larger anomalous TGV's at present or future colliders would
indicate that the new physics responsible for them originates below the
1~TeV scale.  It is therefore likely, though not certain, that the new
physics will first be found directly, rather than through (indirect)
contributions to anomalous TGV's.

There is indirect evidence from constraints on oblique correction parameters
(2-point functions) that anomalous TGV's are indeed $\lsim {\cal
O}(10^{-2})$. The limits obtained from these constraints, however, do depend
strongly on other parameters, such as the Higgs boson and top quark
mass (in the framework where the electroweak symmetry is realized
linearly; see Fig.~1). They also strongly
depend on naturalness arguments which, though compelling,
cannot be considered a proof that large anomalous TGV's do not exist.
Strictly speaking, anomalous TGV's are unconstrained by
the electroweak precision data since the possibility of large cancellations
cannot be excluded. Thus, it is necessary for experiments to search directly
for evidence of anomalous TGV's, even though, in light of our
current theoretical understanding, such experiments will likely yield
null results.

\vglue 0.3cm
{\elevenbf\noindent 3. Measuring $WWV$ and $Z\gamma V$ Couplings in
Collider Experiments}
\vglue 0.2cm
{\elevenit\noindent 3.1 General Overview}
\vglue 0.2cm

In this Section we discuss possibilities to measure the $WWV$ and
$Z\gamma V$ couplings directly in collider experiments. To simplify our
discussion, we assume that $g_1^\gamma=1$ and $g_5^\gamma=g_5^Z=0$ in
the following. As we have mentioned in the Introduction,
electromagnetic gauge invariance requires $g_1^\gamma=1$ and
$g_5^\gamma=0$ for on-shell photons. In contrast to the other couplings,
$g_5^V$, $V=\gamma,\,Z$, violates charge conjugation and parity.
Possibilities to measure $g_5^Z$ in $e^+e^-$ collisions are discussed in
Ref.~\cite{g5}.

At hadron colliders (Tevatron, LHC), di-boson
production offers the best opportunity to probe the $WWV$ and $Z\gamma V$
vertices. The generic set of Feynman
diagrams contributing to di-boson production is shown in Fig.~2. Whereas
$W^+W^-$ production is sensitive to $WW\gamma$ and $WWZ$
couplings, only the $WW\gamma$ ($WWZ$) vertex is tested in
$W^\pm\gamma$ ($W^\pm Z$) production. $Z\gamma V$ couplings are probed in
$pp,\, p\bar p\rightarrow Z\gamma$. In order to reduce the QCD
background, one has to require that at least one of the $W$ and/or $Z$
bosons decays leptonically. In $pp, p\bar p\rightarrow W^+W^-$,
$t\bar t$ production represents an additional background.
$W^\pm\gamma$ and $W^\pm Z$ production are of special
interest due to the presence of amplitude zeros~\cite{WGZERO,WZZERO}.

Electroweak boson pair production at hadron colliders will be discussed
in detail in Section~3.2. We present the general strategy in extracting
information on three vector boson couplings, summarize the current
limits on $WWV$ and $Z\gamma V$ couplings from CDF and D\O, and
investigate the prospects of measuring these couplings in future
Tevatron and LHC experiments. We also discuss possibilities to search
for the amplitude zeros in $W^\pm\gamma$ and $W^\pm Z$ production.
\begin{center}\begin{picture}(400,100)(0,0)
\ArrowLine(10,70)(50,70)
\Vertex(50,70){2}
\ArrowLine(50,70)(50,30)
\Vertex(50,30){2}
\ArrowLine(50,30)(10,30)
\Photon(50,70)(90,70){3}{6}
\Photon(50,30)(90,30){3}{6}
\Text(0,70)[]{\small $q_1$}
\Text(0,30)[]{\small $\bar q_2$}
\Text(100,70)[]{\small $V_1$}
\Text(100,30)[]{\small $V_2$}
\ArrowLine(150,70)(190,70)
\Vertex(190,70){2}
\ArrowLine(190,70)(190,30)
\Vertex(190,30){2}
\ArrowLine(190,30)(150,30)
\Photon(190,70)(230,70){3}{6}
\Photon(190,30)(230,30){3}{6}
\Text(140,70)[]{\small $q_1$}
\Text(140,30)[]{\small $\bar q_2$}
\Text(240,70)[]{\small $V_2$}
\Text(240,30)[]{\small $V_1$}
\ArrowLine(290,70)(320,50)
\Vertex(320,50){2}
\ArrowLine(320,50)(290,30)
\Photon(320,50)(360,50){3}{6}
\Vertex(360,50){2}
\Photon(360,50)(390,70){3}{6}
\Photon(360,50)(390,30){3}{6}
\Text(280,70)[]{\small $q_1$}
\Text(280,30)[]{\small $\bar q_2$}
\Text(400,70)[]{\small $V_1$}
\Text(400,30)[]{\small $V_2$}
\Text(340,60)[]{\small $V$}
\end{picture}\\
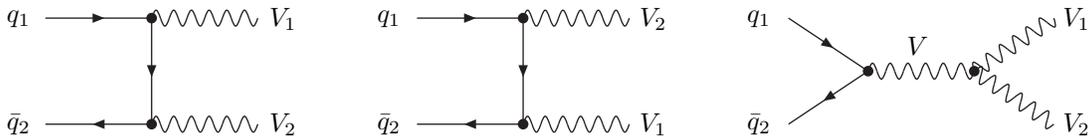
{\sl Figure~2: Generic Feynman diagrams contributing to di-boson
production in hadronic collisions. $V,\,V_1,\,V_2=W,\,\gamma,\, Z$.}
\end{center}

At LEP, $Z\gamma V$ couplings can be tested in single photon production
($e^+e^-\rightarrow\bar\nu\nu\gamma$) and radiative $Z$ decays. Single
photon production, in principle, is also sensitive to the $WW\gamma$
vertex~\cite{GILLES}. $WWZ$ couplings can be probed in the rare decay
$Z\rightarrow Wf\bar f'$~\cite{BH}. In both cases, however, the
sensitivity is not sufficient to compete with the existing limits from CDF
and D\O\ (see below). The constraints on $Z\gamma V$ couplings from
LEP experiments will be discussed in Section~3.3.1.
At LEP~II, $e^+e^-\rightarrow
W^+W^-$ and $e^+e^-\rightarrow Z\gamma$ are the prime reactions to test
$WWV$ and $Z\gamma V$ couplings (see Section~3.3.2). $W$ pair production
at a linear $e^+e^-$ collider with a center of mass energy of 500~GeV or
more [``Next Linear Collider'' (NLC)] will be discussed in
Section~3.3.3. Using
laser backscattering~\cite{LASER}, the NLC can also be operated as a
$e\gamma$ or $\gamma\gamma$ collider, with a center of mass energy of up
to $\sim 80\%$ of that available in the $e^+e^-$ mode, and comparable
luminosity. This opens the
possibility of testing the $WWV$ couplings in processes such
as $e\gamma\rightarrow W\nu$~\cite{ERAN1,shrimp}, or
$\gamma\gamma\rightarrow W^+W^-$~\cite{shrimp,ERAN2} in addition to
$e^+e^-\rightarrow W^+W^-$. $Z\gamma
V$ couplings can be investigated in $Z\gamma$ production and in
$\gamma e\to Ze$~\cite{CHOI}. The NLC
could even be operated as an $e^-e^-$ collider. Possibilities to probe
the three vector boson couplings in $e^-e^-$ collisions have been
explored in Ref.~\cite{FRANK}. The
limits on anomalous $WWV$ couplings expected from reactions
accessible in $e\gamma$, $\gamma\gamma$ and $e^-e^-$ collisions are
similar to those from $e^+e^-\rightarrow W^+W^-$. Alternative $e^+e^-$
processes, such as $e^+e^-\rightarrow
e^+e^-W^+W^-$, $W^+W^-V$ ($V=\gamma,\,Z$), or $\bar\nu\nu Z$ are
significantly less sensitive to three gauge boson couplings than $W$ pair
production. For a summary of these modes see Ref.~\cite{MIYA}. The
sensitivity bounds obtained from $e^+e^-\rightarrow W^+W^-$ are
therefore representative for the limits on anomalous gauge boson
couplings which can be achieved at the NLC.

The $WWV$ couplings can, in principle, also be tested in single $W$ and
$Z$ production at HERA~\cite{EP}. However, in order to
achieve limits which are comparable to the current CDF/D\O\ bounds (see
Section~3.2.2),
integrated luminosities of the order 1~fb$^{-1}$ are needed. Since it
is not expected that those can be achieved within the next few years,
anomalous gauge boson couplings at HERA will not be discussed in this
report.

\vglue 0.2cm
{\elevenit\noindent 3.2 Di-boson Production at Hadron Colliders}
\vglue 0.2cm
{\sl\noindent 3.2.1 Theoretical Background}
\vglue 0.1cm

{}From the phenomenological effective Lagrangian [see Eqs.~(\ref{LeffWWV})
and~(\ref{EQ:ZgammaV})] it is straightforward
to derive cross section formulas for the di-boson production processes,
\begin{eqnarray}
q\bar q\rightarrow W^+W^-,\, Z\gamma,
\label{EQ:REACone}
\end{eqnarray}
and
\begin{eqnarray}
q\bar q'\rightarrow W^\pm\gamma,\, W^\pm Z.
\label{EQ:REACtwo}
\end{eqnarray}
For our subsequent discussion we find it convenient to briefly discuss
the contributions of anomalous couplings to the helicity amplitudes of
the processes listed in~(\ref{EQ:REACone}) and~(\ref{EQ:REACtwo}). In
$q\bar q'\rightarrow W\gamma$, for example, the anomalous contributions
$\Delta{\cal M}_{\lambda^\gamma \lambda^W}$, ($\lambda^\gamma$ and
$\lambda^W$ are the photon and $W$ helicities, respectively) to the
helicity amplitudes are given by~\cite{UBELB1}
\begin{eqnarray}
\Delta{\cal M}_{\pm 0} &=& {e^2 \over \sin\theta_W}
{\sqrt{\hat s} \over 2m_W} \left( \Delta\kappa_\gamma + \lambda_\gamma
\right) \,\textstyle{1\over 2}\,(1\mp \cos\Theta)\>, \\
\Delta {\cal M}_{\pm\pm} &=& {e^2 \over \sin\theta_W}~{1\over 2}\left (
{\hat s \over m^2_W}~ \lambda_\gamma  + \Delta
\kappa_\gamma \right ) \textstyle{1\over \sqrt 2}\,
\sin\Theta\>,
\label{EQ:AMP}
\end{eqnarray}
where $\Theta$ denotes the scattering angle of the photon with respect to
the quark direction, measured in the $W\gamma$ rest frame, and
$\sqrt{\hat s}$ is the invariant mass of the $W$-photon system. Similar
expressions can be derived for the anomalous contributions to the $WZ$,
$W^+W^-$ and $Z\gamma$ helicity amplitudes.

While the SM contribution to the di-boson
amplitudes is bounded from above for fixed scattering angle $\Theta$,
the anomalous contributions rise without
limit as $\hat s$ increases, eventually violating unitarity. This is
the reason the anomalous couplings must show a form
factor behavior at very high energies (see Section~2.3). Anomalous values of
$\lambda_V$, $V=\gamma,\, Z$, are enhanced by $\hat s/m_W^2$ in the
amplitudes
${\cal M}_{\pm\pm}$ for all di-boson production processes. Terms
containing $\Delta\kappa_V$
mainly contribute to ${\cal M}_{\pm 0}$ in $WV$ production and grow
only with $\sqrt{\hat s}/m_W$. In $q\bar q\to W^+W^-$, on the other
hand, the
$\Delta\kappa_V$ term mostly contributes to the (0,0) amplitude and
is enhanced by a factor $\hat s/m_W^2$~\cite{HPZH}. Non-standard values
of $\Delta
g_1^Z$ mostly affect the (0,0) [$(\pm,0)$ and $(0,\pm)$] amplitude in
$WZ$ [$W^+W^-$] production, and are proportional to $\hat s/m_W^2$
[$\sqrt{\hat s}/m_W$]~\cite{HPZH,WZ}. The best limits on
$\Delta\kappa_V$ ($\Delta g_1^Z$) are therefore expected from $q\bar
q\to W^+W^-$ ($q\bar q'\to WZ$). In $Z\gamma$ production, terms
proportional to $h^V_3$ ($h^V_4$) grow like $(\sqrt{\hat
s}/m_Z)^3$ ($(\sqrt{\hat s}/m_Z)^5$)~\cite{UBELB2}.

For large values of the di-boson invariant
mass $\sqrt{\hat s}$, the non-standard contributions to the helicity
amplitudes would dominate, and would suffice to explain differential
distributions
of the photon and the $W/Z$ decay products. Due to the fact that
anomalous couplings only contribute via $s$-channel $W$, $Z$ or photon
exchange, their effects are concentrated in the region of small
vector boson rapidities, and the transverse momentum distribution of the
vector boson should be particularly sensitive to non-standard $WWV$ and
$Z\gamma V$ couplings. This is demonstrated in Fig.~3, where we show the
photon $p_T$ distribution in $p\bar p\to W^+\gamma\to e^+
\nu_e\gamma$, and the $Z$ boson transverse momentum distribution in
$p\bar p\to W^+Z\to\ell_1^+\nu_1\ell_2^+\ell_2^-$, $\ell_{1,2}=e,\,
\mu$, at the Tevatron for the SM and various anomalous $WWV$ couplings.
A dipole form factor (see Section~2.3) with scale $\Lambda_{FF}=1$~TeV
has been assumed.
Only one coupling is assumed to deviate from the SM at a time.
\begin{figure}[t]
\vskip 9.cm
\includegraphics{fig2.ps}
{\sl\noindent Figure~3: The differential cross section for the
transverse momentum  of a) the photon in $p\bar p\to W^+\gamma$, and b)
of the $Z$ boson in $p\bar p\to W^+Z$ at the Tevatron in the SM case
(solid line) and for various anomalous $WWV$ couplings. The cuts imposed
are described in the text.}
\end{figure}
To simulate detector response, the following cuts have been imposed in
Fig.~3:
\begin{eqnarray}
p_T(\gamma)> 10~{\rm GeV}, & \qquad & |\eta(\gamma)|<1,  \nonumber\\
p_T(\ell)> 20~{\rm GeV}, & \qquad & |\eta(\ell)|<2.5,~~\ell=e,\,\mu ,
\label{EQ:WGCUT}\\
p\llap/_T>20~{\rm GeV}, & \qquad & \Delta R(\ell,\ell)>0.4,\nonumber\\
m_T(\ell\gamma;p\llap/_T)>90~{\rm GeV},  &  \qquad &
\Delta R(\gamma,\ell) > 0.7. \nonumber
\end{eqnarray}
Here, $p\llap/_T$ denotes the missing transverse momentum, $\eta$ the
pseudorapidity, $\Delta R=[ (\Delta\phi)^2 + (\Delta\eta)^2]^{1/2}$ the
separation in the pseudorapidity -- azimuthal angle plane, and $m_T$ is
the cluster transverse mass defined by
\begin{eqnarray}
m_T^2 (\ell\gamma;p\llap/_T^{}) & = &
\biggl[ \Bigl( m(\ell\gamma)^2
 + \bigl| \hbox{\bf p}_{T}^{}(\ell\gamma) \bigr|^2 \Bigr)^{1/2}
 + p\llap/_T^{} \biggr]^2 - \bigl|   \hbox{\bf p}_{T}^{}(\ell\gamma)
         + \hbox{\bf p}\llap/_T^{} \bigr|^2 \, ,
\end{eqnarray}
with $m(\ell\gamma)$ being the $\ell\gamma$ invariant mass. The large
lepton photon separation and the $m_T$ cut together strongly suppress
photon radiation from the final state lepton line (radiative $W$
decays)~\cite{UBELB1}.

Information on anomalous $WWV$ and $Z\gamma V$ couplings can be obtained
by comparing the shape of the measured and predicted $p_T$
distribution, provided that the signal is not overwhelmed by background.
If the background is much larger than the SM prediction, limits on
anomalous couplings can still be extracted if a phase space region can
be selected where the effects of non-standard three vector boson
couplings dominate.

Besides di-boson production, radiative $W$ ($Z$) decays are also
sensitive to $WW\gamma$ ($Z\gamma V$) couplings. However, the parton
center of mass energy in these processes is restricted to values around
$\sqrt{\hat s}=m_W$ ($m_Z$), and the expected limits on anomalous
couplings are significantly worse than those obtained from $W\gamma$ and
$Z\gamma$ production where much larger values of $\sqrt{\hat s}$ are
accessible.

\vglue 0.2cm
{\sl\noindent 3.2.2 Di-boson Production at the Tevatron: Current Results
and Future Prospects}
\vglue 0.2cm

Both, the CDF and D\O\ Collaboration have searched for
$W\gamma$~\cite{CDFWG,DOWG},
$Z\gamma$~\cite{CDFZG,DOZG}, $W^+W^-$~\cite{CDFWWWZ,DOWW}, and
$WZ$~\cite{CDFWWWZ} production in the data samples accumulated in
run~1a. CDF has also searched for $W\gamma$ and $Z\gamma$ events in the
data of the 1988 -- 89 run~\cite{CDFOLD}. For a recent summary of
electroweak boson pair production results from CDF and D\O\ see
Ref.~\cite{Steve}.

CDF (D\O) extract $W\gamma/Z\gamma$ data samples from inclusive
$e/\mu$ channel $W/Z$ samples by requiring an isolated photon
in a fiducial region of their central (central + endcap) electromagnetic
(EM) calorimeters with $E_{T}(\gamma) \ge 7~(10) ~{\rm GeV}$.
A minimum lepton $-$ photon angular separation of
$\Delta R(\ell \gamma) > 0.7$ suppresses final-state QED bremsstrahlung.
To reduce the QCD background from $W/Z$+jets production, excess
calorimeter transverse energy, $E_T$,
within a cone of $\Delta R = 0.4$ centered on the photon was required to be
less than 15\% (10\%) of the photon $E_T$. CDF also required the sum of the
transverse momenta of all charged tracks within this cone to be less
than $2~{\rm GeV/c},$
and also rejected events with a track pointing directly at the EM cluster.
Both experiments required transverse/longitudinal EM shower
development consistent with a single photon. The
selection criteria yield~25 (23) $W\gamma$ and 8 (6) $Z\gamma$
candidate events for CDF (D\O).

The level of $W/Z$+jet background, where a jet ``fakes'' an isolated
photon, in each of the $W\gamma/Z\gamma$ data samples
is determined by use of QCD jet data samples to obtain
a jet misidentification probability ${\cal P}_{j \rightarrow \gamma}(E_T)$.
For the photon selection criteria used by CDF,
${\cal P}_{j \rightarrow \gamma}(E_T = 9~{\rm GeV}) \sim 8\times 10^{-4},$
decreasing exponentially to
${\cal P}_{j \rightarrow \gamma}(E_T = 25~{\rm GeV}) \sim 10^{-4},$ whereas
for the photon selection criteria used by D\O,
${\cal P}_{j \rightarrow \gamma}(E_T) \sim 4\times 10^{-4}$ ($6\times
10^{-4}$) in the central (endcap) calorimeter, and varies only slowly with
$E_T$. The jet fragmentation probability distribution was then convoluted
with the jet $E_T$ spectrum in each of the inclusive $W/Z$ data samples.
The $Z\gamma$ background in the $W\gamma$ data arising from non-observation
of one of the $Z$ decay leptons is estimated from Monte Carlo simulations.
The contributions to $W\gamma$ and $Z\gamma$ production from $W/Z$
decays into $\tau$ leptons are also estimated from MC simulations and
found to be small.

The SM cross sections for $W^+W^-$, $W^\pm Z$ and $ZZ$
production\footnote{$ZZ$ production is, in principle, sensitive to
$ZZV$, $V=\gamma,\, Z$ couplings, which vanish in the SM at tree
level~\cite{HPZH}. We will not discuss the $ZZV$ couplings accessible
in $ZZ$ production in this report.} at the
Tevatron, including NLO QCD corrections~\cite{JIM}, are 9.5~pb,
2.5~pb and 1.4~pb, respectively. Decay modes where one of the weak
bosons decays hadronically have significantly larger branching ratios
than all leptonic decays:
\begin{eqnarray}
Br(WW\to e\nu_e e\nu_e,\mu\nu_\mu\mu\nu_\mu)=2.4\%, & \qquad & Br(WW\to
e\nu_e\mu\nu_\mu)=2.4\%, \\
Br(WZ\to\ell_1\nu_1\ell_2^+\ell_2^-)=1.5\%, &\qquad & \ell_{1,2}=e,\,\mu
, \\
Br(WW\to\ell\nu jj)=29\%, & \qquad & \ell=e,\,\mu , \\
Br(WZ\to\ell\nu jj)=15\%, & \qquad & Br(WZ\to jj\ell^+\ell^-)=4.5\%, \\
BR(ZZ\to\ell^+\ell^- jj)=9.4\%, &\qquad &
Br(ZZ\to\ell_1^+\ell_1^-\ell_2^+\ell_2^-)=0.4\%.
\end{eqnarray}
Due to the larger cross section and branching ratio, the
$\ell\nu jj$ final state is completely dominated by $W^+W^-$ production.
$W^\pm Z$ and $ZZ$ production contribute approximately equally to the
$\ell^+\ell^- jj$ final state.
All semihadronic channels suffer from a large $W/Z+$~jets background.
$t\bar t$ production contributes non-negligibly to the background for
$W^+W^-$ production. In contrast, the $\ell_1\nu_1\ell_2^+\ell_2^-$ final
state is relatively background free.

$W^+W^-$ and $WZ$ data samples are also extracted from inclusive $e/\mu$
$W/Z$ data. CDF has analyzed the $WW, WZ \rightarrow \ell \nu jj$ and
$ZW \rightarrow \ell^+\ell^-jj$ ($\ell = e$ or $\mu$) channels using
standard $W/Z$ lepton selection cuts, and requiring $60~{\rm GeV/c^2} <
m(jj) < 110~{\rm GeV/c^2}$.
For leptonic $W~(Z)$ events, CDF eliminates $W/Z$+jets background events
by requiring
$p_T(jj) > 130 ~(100)~{\rm GeV/c},$ which also eliminates the
SM signal but retains good sensitivity for non-zero $WWV$ anomalous
couplings. One event passes the cuts in the $\ell\nu jj$ channel. In
the $\ell^+\ell^-jj$ channel no events survive. A clean candidate event
for $p\bar p\to W^+Z\to e^+\nu_e e^+e^-$ has also been observed in the
CDF data set~\cite{CDFWWWZ}. D\O\ has analyzed the $WW \rightarrow
\ell_1\nu_1\ell_2\nu_2$, $\ell_{1,2}=e,\,\mu$, channels using
standard lepton cuts for selection of $W$ pairs. The $Z$ mass region in
the $ee$ channel, $77~{\rm GeV/c^2} < m(ee) < 105~{\rm
GeV/c^2}$, is excluded. To suppress the $Z\to\mu^+\mu^-$ background, a
cut of $E\llap/^\eta_T>30$~GeV is imposed, where $E\llap/^\eta_T$ is the
projection of the missing $E_T$ vector onto the bisector of the decay
angles of the two muons.
To reduce the $t\bar t$ background, the
total hadronic transverse energy in the event is required to be less
than 40~GeV. Backgrounds from $Z$ decay and fake
electrons are estimated from data and MC simulations. One $ee\nu\nu$ and
one $e\mu\nu\nu$ event pass all cuts.

SM and anomalous coupling predictions for the $W\gamma$ and $Z\gamma$
processes are obtained using the event generators of Ref.~\cite{UBELB1}
and~\cite{UBELB2}, and detailed detector simulations. $\rm MRSD-'$
structure functions~\cite{MRS} are
used for event generation as they best match the recent $W$ lepton
asymmetry measurements from CDF~\cite{ASYMMETRY}.
SM and anomalous coupling predictions for $W^+W^-$ and $WZ$ production
are obtained using the event generator of Ref.~\cite{HWZ}
and MC detector simulations. Presently, a complete calculation of the
di-boson transverse momentum distribution, including soft gluon
resummation effects, does not exist, except for the $ZZ$
case~\cite{HMO}. Higher order QCD corrections are
therefore approximated in the experimental analysis by a $k$-factor and
by smearing the transverse momentum of the di-boson system according to
the experimentally determined $W/Z$ boson $p_T$ spectrum.

Direct experimental limits on $WW\gamma$ and $Z\gamma V$
anomalous couplings for the $W\gamma/Z\gamma$ processes are obtained
via binned maximum likelihood fits to the $E_T(\gamma)$ distribution.
The observed $E_T(\gamma)$ distribution is compared to the sum of
expected signal plus background(s) prediction, calculating the Poisson
likelihood that this sum would fluctuate to the observed number of
events in each $E_T$ bin, and convoluting with a Gaussian distribution
to take into account systematic uncertainties associated with backgrounds,
luminosity normalization, structure function choice, $Q^2$-scale and
uncertainties in the shape of the $p_T(W\gamma/Z\gamma)$ distribution,
efficiencies, {\it etc}. The 95\% CL
CDF~\cite{CDFWG} and D\O~\cite{DOWG} limits on anomalous
$WW\gamma$ couplings from $W\gamma$ production are shown in Fig.~4a. The
bounds on $\Delta\kappa_\gamma^0$ and $\lambda_\gamma^0$ extracted by
D\O\ (solid curve) are about 20\% better than those obtained by CDF
(short dashed curve). For
comparison, we have also included the limits obtained by UA2~\cite{UA2},
and CDF from the 1988-89 data~\cite{CDFOLD}. Due to the smaller center
of mass energy ($\sqrt{s}=630$~GeV), the correlations between the two
couplings at the CERN $p\bar p$ collider are much more pronounced than
at Tevatron energies. The bounds obtained from the 1992-93 data have
been obtained using a dipole form factor with scale
$\Lambda_{FF}=1.5$~TeV. The CDF limits from the 1988-89 data are for
$\Lambda_{FF}=1$~TeV.
\begin{figure}[t]
\vskip 7.5cm
\includegraphics{fig3a.ps}
\includegraphics{fig3b.ps}
{\sl\noindent Figure~4: Present limits on anomalous $WWV$ couplings from
hadron collider experiments.}
\end{figure}

CDF extracts direct experimental limits on $WW\gamma$ and $WWZ$
anomalous couplings from the $\ell\nu jj$ and $\ell^+\ell^- jj$ final states
via comparison of observed events to the expected signal within cuts,
including systematic uncertainties due to luminosity normalization,
jet energy scale and resolution, structure function choice and
higher order QCD corrections, {\it etc}.
D\O\ extracts direct experimental limits on $WWV$
anomalous couplings from the $WW\to \ell_1\nu_1\ell_2\nu_2$ mode via
comparison of their 95\% CL upper limit of $\sigma(WW)_{expt} \! < \!
91~{\rm pb}$ with $\sigma(WW)_{pred}$ as a function of anomalous couplings.

The limits obtained from $W^+W^-\to \ell_1\nu_1\ell_2\nu_2$ and $WW,\,
WZ\to\ell\nu jj$ are summarized and
compared to those obtained from $W\gamma$ production in Fig.~4b. In
extracting limits on non-standard $WWV$ couplings from $W$ pair
production, CDF (D\O) assumed a dipole form factor with scale
$\Lambda_{FF}=1.5$~TeV (0.9~TeV), $\Delta\kappa_\gamma^0=\Delta\kappa_Z^0$,
$\lambda_\gamma^0=\lambda_Z^0$, and $\Delta g_1^Z=0$. Due to the
selection of a phase space region which is particularly sensitive to
$WWV$ couplings and the larger branching ratios for $WW,\,WZ\to\ell\nu jj$,
the bounds obtained from the semihadronic $WW$ and $WZ$ final states are
significantly stronger than those found from analyzing the $WW\to
\ell_1\nu_1\ell_2\nu_2$ channel. In Section~3.2.1 we have mentioned
that the contributions to the $W^+W^-$ helicity amplitudes proportional to
$\Delta\kappa_V$ grow like $\hat s/m_W^2$ whereas the
$\Delta\kappa_\gamma$ terms in the $W\gamma$ amplitudes are proportional
to $\sqrt{\hat s}/m_W$. In contrast, the $\lambda_V$ terms always grow
like $\hat s/m_W^2$. This explains why the limit on $\Delta\kappa_V^0$
obtained from the semihadronic $WW$ and $WZ$ final states is
significantly better than that found from $p\bar p\to W\gamma$ while the
bounds on $\lambda_V^0$ from $WW,\,WZ\to\ell\nu jj$ and $W\gamma$
production are almost identical.

Limits on $\Delta\kappa_\gamma^0$ and $\lambda_\gamma^0$ extracted from
$W\gamma$ production have the advantage of being independent of
assumptions about the $WWZ$ vertex. Similarly, information on the $WWZ$
couplings, independent from assumptions on the $WW\gamma$ couplings, can
be obtained from $WZ$ production. From the $WZ\to jj\ell^+\ell^-$
channel, CDF finds~\cite{CDFWWWZ}
$-8.6<\Delta\kappa_Z^0<9.0$ for $\Delta g_1^{Z0}=\lambda_Z^0=0$
and $-1.7<\lambda_Z^0<1.7$ for $\Delta g_1^{Z0}=\Delta\kappa_Z^0=0$.
These limits were obtained for a form factor scale $\Lambda_{FF}=1.
5$~TeV. The $ZZ\to jj\ell^+\ell^-$ cross section was assumed to be given
by the SM prediction.

\begin{figure}[t]
\vskip 7.5cm
\includegraphics{fig4a.ps}
\includegraphics{fig4b.ps}
{\sl\noindent Figure~5: Comparison of current experimental bounds on
$WWV$ couplings and limits obtained from $S$-matrix unitarity for a
dipole form factor.}
\end{figure}
In Section~2.3 we have seen that constraints from $S$-matrix unitarity
severely restrict the values of the low energy anomalous couplings
allowed. For sufficiently small values of the form factor
scale, the experimental limits on non-standard three vector boson
couplings are substantially better than those found from $S$-matrix
unitarity [see Eqs.~(\ref{EQ:unitbf}) --~(\ref{EQ:unitbl})]. However, for
$\Lambda_{FF}\gg m_W$, the unitarity
bounds decrease like $1/\Lambda_{FF}^n$, with $n=1,2$ for the $WWV$
couplings, and $n=3,5$ for the $Z\gamma V$ couplings whereas the
experimental
limits depend less sensitively on $\Lambda_{FF}$~\cite{CDFOLD}.
This implies that for sufficiently large form factor
scales unitarity bounds eventually will be stronger than the limits
extracted from experimental data. In Fig.~5a we compare the current
experimental limits on $WW\gamma$ couplings from $W\gamma$
production with the bounds derived from unitarity
for $\Lambda_{FF}=1.5$~TeV. In Fig.~5b a similar comparison is carried
out for $WW/WZ\to\ell\nu jj$ with $\Lambda_{FF}=1.5$~TeV, and
$WW\to\ell_1\nu_1\ell_2\nu_2$ with $\Lambda_{FF}=0.9$~TeV. These values
of $\Lambda_{FF}$
were chosen just large enough that the unitarity bounds would approach
the experimental limits. One concludes that the maximum scale which can
be probed with the current experimental data on $W\gamma$, $WW$ and
$WZ$ production is of order 1.5~--~2~TeV.

\begin{figure}
\vskip 7.5cm
\includegraphics{fig5.ps}
{\sl\noindent Figure~6: Present limits on anomalous $ZZ\gamma$
couplings from
hadron collider experiments, and constraints from $S$-matrix unitarity.}
\end{figure}
The current CDF~\cite{CDFZG} and D\O~\cite{DOZG} 95\% CL limit
contours for anomalous $ZZ\gamma$ couplings are shown in
Fig.~6, together with the constraints from $S$-matrix unitarity. The
limit contours for $Z\gamma\gamma$ couplings are similar. For
completeness, we have also included the CDF result from the 1988-89
run~\cite{CDFOLD}. The D\O\ limits on $h_{30}^V$ and $h_{40}^V$ are
about 30\% more stringent than those obtained by CDF.
In order to derive these limits, generalized dipole form factors
with $\Lambda_{FF}=0.5$~TeV, and powers $n=3$ ($n=4$) for $h_3^V$
($h_4^V$), are assumed (see Section~2.3).
Since the anomalous contributions to the $Z\gamma$ helicity amplitudes
grow faster with energy than those in $W\gamma$ production, the
experimental limits on $h_{30}^V$ and $h_{40}^V$ depend rather
sensitively on the form factor scale chosen. The maximum form factor
scale which can be probed in $Z\gamma$ production with present
experimental data is $\Lambda_{FF}\approx 500$~GeV.

Table~1 summarizes the current results on anomalous $WWV$ and $Z\gamma V$
couplings from hadron colliders. With the limited statistics of di-boson
events currently available, deviations from the SM cross section have
to be large at least in some regions of phase space in
order to lead to an observable effect. The best direct limits on
$\Delta\kappa_V^0$ are currently obtained from the $\ell\nu jj$ final state.
$W\gamma$ production results in somewhat better bounds on
$\lambda_\gamma^0$ than $p\bar p\to WW,\, WZ\to\ell\nu jj$.
So far, no attempt has been made to combine the limits of CDF and D\O\
and/or from different channels.

During the current data taking period (run~1b) at the Tevatron, one hopes
to collect an integrated luminosity of about 100~pb$^{-1}$ per
experiment. For the Main Injector Era,
integrated luminosities of the order of 1~fb$^{-1}$ are
envisioned~\cite{MI}. The first run with the Main Injector is currently
planned for the period of 1998 -- 2003. Through further upgrades of the
Tevatron
accelerator complex, an additional factor~10 in luminosity may be gained
(TeV*). The substantial increase in integrated luminosity will make it
possible to test the $WWV$ and $Z\gamma V$ vertices with much greater
precision than in current experiments. In Fig.~7 we show the 95\% CL
limits on anomalous $WW\gamma$ and $ZZ\gamma$ couplings expected for CDF
from $W\gamma$ and $Z\gamma$ production at the Tevatron
($\sqrt{s}=2$~TeV) for
1~fb$^{-1}$ and 10~fb$^{-1}$. Here, and in all subsequent sensitivity
plots, we assume that no deviation from the SM prediction is observed in
future experiments. To derive bounds on non-standard $WWV$ couplings a
dipole form factor is assumed. For the $Z\gamma V$ couplings we use form
factor powers of $n=3$ ($h_3^V$) and $n=4$ ($h_4^V$).
\begin{table}
\caption{95\% CL limits on anomalous $WWV$, $V=\gamma, \, Z$, and
$ZZ\gamma$ couplings from CDF and D\O. Only one of the independent
couplings is allowed to deviate from the SM at a time. The bounds
obtained for $Z\gamma\gamma$ couplings are very similar to those derived
for the $ZZ\gamma$ couplings and are therefore not shown.}
\begin{center}
\begin{tabular}{|ccc|}
\hline
& & \\[-3.mm]
experiment & channel & limit \\[1.mm] \hline
& & \\[-3.mm]
CDF & $p\bar p\rightarrow W^\pm\gamma\rightarrow\ell^\pm\nu\gamma$ &
$-2.3<\Delta\kappa_\gamma^0<2.2$ \\[1.mm]
 & $\ell=e,\,\mu$ & $-0.7<\lambda_\gamma^0<0.7$ \\[1.mm] \hline
& & \\[-3.mm]
D\O\ & $p\bar p\rightarrow W^\pm\gamma\rightarrow\ell^\pm\nu\gamma$ &
$-1.6<\Delta\kappa_\gamma^0<1.8$ \\[1.mm]
 & $\ell=e,\,\mu$ & $-0.6<\lambda_\gamma^0<0.6$ \\[1.mm] \hline
& & \\[-3.mm]
CDF& $p\bar p\rightarrow W^\pm Z\rightarrow\ell^+\ell^-jj$ & $-8.6 < \Delta
\kappa_Z^0<9.0$ \\[1.mm]
 & $\ell=e,\,\mu$ & $-1.7<\lambda_Z^0<1.7$ \\[1.mm] \hline
& & \\[-3.mm]
CDF & $p\bar p\rightarrow W^+W^-,\, W^\pm Z\rightarrow \ell^\pm\nu jj$ &
$-1.0 <\Delta\kappa_V^0<1.1$ \\[1.mm]
 & $\ell=e,\,\mu$, $\kappa_\gamma=\kappa_Z$, $\lambda_\gamma=\lambda_Z$
 & $-0.8<\lambda_V^0<0.8$ \\[1.mm] \hline
& & \\[-3.mm]
D\O\ & $p\bar p\rightarrow W^+W^-\rightarrow \ell_1\nu_1\ell_2\nu_2$ &
$-2.6 < \Delta\kappa_V^0 < 2.8$ \\[1.mm]
 & $\ell_{1,2}=e,\,\mu$, $\kappa_\gamma=\kappa_Z$,
$\lambda_\gamma=\lambda_Z$ & $-2.2 < \lambda_V^0 < 2.2$ \\[1.mm] \hline
& & \\[-3.mm]
CDF & $p\bar p\rightarrow Z\gamma\rightarrow\ell^+\ell^-\gamma$ &
$-3.0<h^Z_{30}<2.9$ \\[1.mm]
 & $\ell=e,\,\mu$, $\Lambda_{FF}=0.5$~TeV & $-0.7<h^Z_{40}<0.7$
\\[1.mm] \hline
& & \\[-3.mm]
D\O\ & $p\bar p\rightarrow Z\gamma\rightarrow\ell^+\ell^-\gamma$ &
$-1.9<h^Z_{30}<1.8$ \\[1.mm]
 & $\ell=e,\,\mu$, $\Lambda_{FF}=0.5$~TeV & $-0.5<h^Z_{40}<0.5$
\\[1.mm] \hline
\end{tabular}
\end{center}
\end{table}
The curves shown in Fig.~7 are obtained
from a binned likelihood fit of the photon $E_T$ distribution.
In the $Z\gamma$ case we also show the
constraint from unitarity for $\Lambda_{FF}=1.5$~TeV. The expected
experimental limits are calculated for the same value of $\Lambda_{FF}$.
\begin{figure}[t]
\vskip 7.5cm
\includegraphics{fig6a.ps}
\includegraphics{fig6b.ps}
{\sl\noindent Figure~7: Projected 95\% CL sensitivity limits for a)
$WW\gamma$
couplings from $W\gamma$ production and b) $ZZ\gamma$ couplings from
$Z\gamma$ production at
the Tevatron for integrated luminosities of 1~fb$^{-1}$ and 10~fb$^{-1}$.}
\end{figure}
The limits on $Z\gamma\gamma$ couplings are very similar to those found for
$ZZ\gamma$ couplings and are therefore not shown. Only $W\to e\nu_e$
and $Z\to e^+e^-$ decays are taken into account in our analysis.
Electrons are required to have $|\eta|<3.6$, with at
least one electron in the central region of the detector ($|\eta|<1.
0$). A pseudorapidity cut of $|\eta(\gamma)|<2.4$ is imposed on photons.
The acceptances are
calculated using the following transverse energy and separation cuts:
\begin{eqnarray}
E_T(e) > 25~{\rm GeV,} & \qquad & E\llap/_T > 25~{\rm GeV,}
\label{EQ:WGSIM1}\\
E_T(\gamma) > 10~{\rm GeV,} &\qquad & \Delta R(e,\gamma) > 0.7.
\label{EQ:WGSIM2}
\end{eqnarray}
In addition, a cut on the transverse $W$ mass of $m_T^W>50$~GeV and a
cluster transverse mass cut of $m_T(e\gamma;E\llap/_T)>90$~GeV were
imposed in the $W\gamma$ case. For $Z\gamma$ production, we require
$m(e^+e^-\gamma) > 100$~GeV and $m(e^+e^-) > 70$~GeV. The efficiencies
for electron and photon
identification were taken from the current CDF analysis, as well as the
probability for a jet to fake a photon, ${\cal P}_{j \rightarrow
\gamma}(E_T)$. The systematic uncertainty from
the integrated luminosity, parton densities, and higher order QCD
corrections was assumed to be 5\%. From Fig.~7a (7b) one observes that
the current limits on anomalous gauge boson couplings can be improved by
about a factor 5~--~15 (10~--~100) in $W\gamma$ ($Z\gamma$) production
in the Main Injector Era. An additional factor 10 in integrated
luminosity leads to roughly a factor~2 improvement in the sensitivities
which can be achieved. The maximum form factor scale which
can be probed in $Z\gamma$ production with 1~fb$^{-1}$ (10~fb$^{-1}$)
is about a factor~2.6 (3) larger than
that accessible with the current data. The limit contours shown in
Fig.~7 can be improved by about 20~--~40\% if $W\to\mu\nu$ and
$Z\to\mu^+\mu^-$ decays are included in the analysis.

The bounds on $Z\gamma V$ couplings could be further improved by
analyzing the reaction $p\bar p\to Z\gamma\to\bar\nu\nu\gamma$. Here the
signal consists of a single high $p_T$ photon accompanied by a large
amount of missing transverse energy. Compared to the charged lepton
decay mode of the $Z$ boson, the decay
$Z\to\bar\nu\nu$ offers potential advantages. Due to
the larger $Z\to\bar\nu\nu$ branching ratio, the differential cross
section is about a factor~3 larger than that for $q\bar q\to
e^+e^-\gamma$ and $q\bar q\to\mu^+\mu^-\gamma$ combined. Furthermore,
final state bremsstrahlung and timelike virtual photon diagrams do not
contribute to the $\bar\nu\nu\gamma$ final state. On the other hand,
there are several potentially serious background processes which
contribute to $p\bar p\to\gamma p\llap/_T$, but not to the
$\ell^+\ell^-\gamma$ final state. The two most important background
processes are prompt photon
production, $p\bar p\to\gamma j$, with the jet rapidity outside the
range covered by the detector and thus ``faking'' missing transverse
momentum, and two jet production where one of the jets is misidentified
as a photon while the other disappears through the beam hole. A parton
level simulation of the $\gamma j$ and $jj$ backgrounds in
$p\bar p\to\gamma p\llap/_T$ suggests~\cite{UBELB2} that those
backgrounds can
be eliminated by requiring a sufficiently large transverse
momentum for the photon.

To estimate the sensitivity of $W^+W^-,\,W^\pm Z\to\ell\nu jj$ and
$WZ\to\ell^+\ell^- jj$, $\ell=e,\,\mu$, to non-standard $WWV$ couplings
in future Tevatron experiments, we require charged leptons to have
$E_T>20$~GeV and $|\eta(\ell)|<2$, and impose a missing transverse
energy cut
of 20~GeV. The two leading jets are required to have $E_T(j)>30$~GeV
and $60~{\rm GeV}<m(jj)<110~{\rm GeV}$.
Events containing an extra jet with $E_T>50$~GeV are vetoed in order to
suppress the top quark background and to reduce the effect of QCD
corrections~\cite{JIM,UJH}. To suppress
the $W/Z+$~jets background, a cut on the transverse momentum of
the jet pair is imposed, similar to the requirement in the current CDF
analysis. The value of the $p_T(jj)$ cut varies with the integrated
luminosity assumed:
\begin{eqnarray}
p_T(jj) > 150~{\rm GeV} &\qquad & {\rm for}\hskip 0.2cm \int\!{\cal
L}dt=100~{\rm pb}^{-1},\\
p_T(jj) > 200~{\rm GeV} & \qquad & {\rm for}\hskip 0.2cm  \int\!{\cal
L}dt=1~{\rm fb}^{-1},\\
p_T(jj) > 250~{\rm GeV} & \qquad & {\rm for}\hskip 0.2cm  \int\!{\cal
L}dt=10~{\rm fb}^{-1}.
\end{eqnarray}
The number of signal events expected is calculated using the event
generator of Ref.~\cite{HWZ}. The trigger and particle identification
efficiencies are assumed to be the same as in the current CDF data
analysis. To estimate the $t\bar t$ and $W/Z+$~jets background, ISAJET
and VECBOS~\cite{WALTER} are used. The top quark mass is taken
to be $m_t=170$~GeV.

Confidence levels are obtained by counting events above the $p_T(jj)$
cut. The resulting 95\% CL contours at $\sqrt{s}=1.8$~TeV for integrated
luminosities of
100~pb$^{-1}$, 1~fb$^{-1}$ and 10~fb$^{-1}$ are shown in Fig.~8a.
To calculate the sensitivity limits in Fig.~8a, we have assumed a form
factor scale of $\Lambda_{FF}=2$~TeV and the effective Lagrangian
scenario of Section~2.2 where the $SU(2)_L\times U(1)_Y$ symmetry is
linearly realized with $f_B=f_W$ (``HISZ
scenario''~\cite{HISZ}), which reduces
the number of independent $WWV$ couplings from five to two. Choosing
$\Delta\kappa_\gamma$ and $\lambda_\gamma$ as independent parameters,
the $WWZ$ couplings are then given by [see Eqs.~(\ref{kapanom})
--~(\ref{kapanomd})]:
\begin{eqnarray}
\Delta g_1^Z & = & {1\over 2\cos^2\theta_W}\,\Delta\kappa_\gamma ,
\label{EQ:HISZ1} \\[1.mm]
\Delta\kappa_Z & = & {1\over 2}\,(1-\tan^2\theta_W)\,\Delta\kappa_\gamma
, \label{EQ:HISZ2}
\\[1.mm]
\lambda_Z & = & \lambda_\gamma .
\label{EQ:HISZ3}
\end{eqnarray}
The sensitivity limits depend only marginally on the value of
$\Lambda_{FF}$ assumed.
\begin{figure}
\vskip 7.5cm
\includegraphics{fig7a.ps}
\includegraphics{fig7b.ps}
{\sl\noindent Figure~8: Expected 95\% CL sensitivity limits for the $WWV$
couplings in the HISZ scenario [see Eqs.~(\ref{EQ:HISZ1}) --
(\ref{EQ:HISZ3})] a) from $p\bar p\to WW,WZ\to\ell\nu jj$ and
$\ell^+\ell^- jj$, and b) from $p\bar p\to
W^\pm Z\to\ell_1^\pm\nu_1\ell_2^+\ell_2^-$ at the Tevatron.}
\end{figure}
The bounds obtained in this scenario are compared in Table~2 with those
derived for different relations between the $WWV$ couplings.
\begin{table}
\caption{95\% CL limits on anomalous $WWV$, $V=\gamma,\, Z$ from $p\bar
p\to WW,\,WZ\to\ell\nu jj$ and $\ell^+\ell^- jj$ at
$\protect{\sqrt{s}=1.8}$~TeV for $\protect{\int\!{\cal
L}dt=1}$~fb$\protect{^{-1}}$ and $\protect{\int\!{\cal
L}dt=10}$~fb$\protect{^{-1}}$. Only one coupling at a time is varied,
except for the dependencies noted.}
\begin{center}
\begin{tabular}{|ccc|}
\hline
& & \\[-3.mm]
dependent couplings & limit & limit\\[1.mm]
 & $\int\!{\cal L}dt=1$~fb$^{-1}$ & $\int\!{\cal L}dt=10$~fb$^{-1}$
 \\[1.mm]\hline
& & \\[-3.mm]
Eqs.~(\ref{EQ:HISZ1}) and (\ref{EQ:HISZ2}) &
$-0.31 <\Delta\kappa_\gamma^0<0.41$ & $-0.17 <\Delta\kappa_\gamma^0<0.
24$ \\[2.mm]
$\lambda_\gamma=\lambda_Z$ &
$-0.19<\lambda_\gamma^0<0.19$ & $-0.10<\lambda_\gamma^0<0.11$ \\[2.mm]
$\Delta\kappa_\gamma=\Delta\kappa_Z$ &
$-0.23 <\Delta\kappa_V^0<0.29$ & $-0.12 <\Delta\kappa_V^0<0.16$ \\[2.mm]
 -- & $-0.35<\Delta g_1^{Z0}<0.43$ & $-0.19<\Delta g_1^{Z0}<0.25$ \\[2.mm]
 -- & $-0.30<\Delta\kappa_Z^0<0.37$ & $-0.16<\Delta\kappa_Z^0<0.20$ \\[2.mm]
 -- & $-0.22<\lambda_Z^0<0.22$ & $-0.11<\lambda_Z^0<0.11$ \\[2.mm]
 -- & $-0.56<\lambda_\gamma^0<0.56$ & $-0.28<\lambda_\gamma^0<0.29$ \\[2.mm]
\hline
\end{tabular}
\end{center}
\end{table}
The sensitivity limits found in the HISZ scenario are seen to be
representative. If
the Tevatron center of mass energy can be increased to 2~TeV the results
shown in Fig.~8a and Table~2 improve by a few per cent.

For integrated luminosities $\ge 1$~fb$^{-1}$, $WW$ and $WZ$ production
with all leptonic
decays can also be used to constrain the $WWV$ vertices. In contrast to
the semihadronic $WW,\, WZ$ production channels, double leptonic $WZ$ decays
are relatively background free and thus provide an excellent testing
ground for non-standard $WWZ$ couplings. Using a recent
calculation of $W^\pm Z$ production which includes NLO QCD
corrections~\cite{UJH}, sensitivity limits for the $p\bar p\to W^\pm
Z\to\ell^\pm_1\nu_1\ell_2^+\ell_2^-$, $\ell_{1,2}=e,\,\mu$, channel
were estimated. No full
detector simulation was carried out, however, lepton identification cuts
of $p_T(\ell_{1,2})>20$~GeV and $|\eta(\ell_{1,2})|<2.5$, and a missing
$p_T$ cut of 20~GeV have been imposed to roughly simulate detector
response. Particle momenta are smeared according to the resolution of
the CDF detector. The 95\% CL limit contours for $\sqrt{s}=1.8$~TeV and
$\Lambda_{FF}=1$~TeV, obtained from a $\chi^2$ fit to the
$p_T(Z)$ distribution are displayed in Fig.~8b. Here we have again
assumed the relations of Eqs.~(\ref{EQ:HISZ1}) -- (\ref{EQ:HISZ3})
for $WW\gamma$ and $WWZ$ couplings. If the center of mass energy of the
Tevatron can be increased to 2~TeV, slightly better limits can be
obtained. For $\int\!{\cal L}dt=1$~fb$^{-1}$,
the small number of $\ell^\pm_1\nu_1\ell_2^+\ell_2^-$ events severely
limits the sensitivity, and the limits obtained from $WW,\,WZ\to\ell\nu
jj$ and $\ell^+\ell^- jj$ are significantly better than those from
double leptonic $WZ$ decays for most of the parameter space. For
10~fb$^{-1}$, the non-negligible
background starts to negatively influence the semihadronic channels, and
double leptonic and $WW,\, WZ\to\ell\nu jj$ and $\ell^+\ell^- jj$ final
states yield comparable results. In contrast to double leptonic $WZ$
decays, the $WW\to\ell_1\nu_1\ell_2\nu_2$ final
states are plagued by background from $t\bar t$
production, and thus were not studied in detail. The contour limits
shown in Figs.~7a and~8 depend only marginally on
the form factor scale assumed; only the limits on the $Z\gamma V$
couplings are more sensitive to the value of $\Lambda_{FF}$ chosen.

The expected sensitivity bounds from future Tevatron experiments,
varying only one of the independent couplings at a time, are
summarized in Table~3.
\begin{table}
\caption{Expected 95\% CL limits on anomalous $WWV$, $V=\gamma,
\, Z$, and $ZZ\gamma$ couplings from future Tevatron experiments. Only
one of the independent couplings is assumed to deviate from the SM at
a time. The limits found
for $Z\gamma\gamma$ couplings are very similar to those obtained for
$h_3^Z$ and $h_4^Z$.}
\begin{center}
\begin{tabular}{|ccc|}
\hline
& & \\[-3.mm]
channel & $\!\!\!\!\!\!\!$ limit & limit\\[1.mm]
 & $\!\!\!\!\!\!\!$ $\int\!{\cal L}dt=1$~fb$^{-1}$ & $\int\!{\cal
L}dt=10$~fb$^{-1}$  \\[1.mm] \hline
& & \\[-3.mm]
$p\bar p\to W^\pm\gamma\rightarrow e^\pm\nu\gamma$ &
$\!\!\!\!\!\!\!-0.38<\Delta\kappa_\gamma^0<0.38$ &
$\!-0.21<\Delta\kappa_\gamma^0<0.21$ \\[1.mm]
$\sqrt{s}=2$~TeV & $\!\!\!\!\!\!\!-0.12<\lambda_\gamma^0<0.12$ &
$\!-0.057<\lambda_\gamma^0<0.057$ \\[1.mm] \hline
& & \\[-3.mm]
$p\bar p\to W^+W^-,\, W^\pm Z\rightarrow \ell^\pm\nu jj,\,\ell^+\ell^-
jj$ &
$\!\!\!\!\!\!\!-0.31 <\Delta\kappa_\gamma^0<0.41$ & $\!-0.17
<\Delta\kappa_\gamma^0<0.24$ \\[1.mm]
$\ell=e,\,\mu$, HISZ scenario &
$\!\!\!\!\!\!\!-0.19<\lambda_\gamma^0<0.19$ & $\!-0.10<\lambda_\gamma^0<0.
11$ \\[1.mm] \hline
& & \\[-3.mm]
$p\bar p\to W^\pm Z\to\ell_1^\pm\nu_1\ell_2^+\ell_2^-$ & $\!\!\!\!\!\!\!-0.
26<\Delta\kappa_\gamma^0<0.70$ & $\!-0.09 < \Delta\kappa_\gamma^0<0.32$
\\[1.mm]
$\ell_{1,2}=e,\,\mu$, HISZ scenario &
$\!\!\!\!\!\!\!-0.24<\lambda_\gamma^0<0.32$ & $\!-0.10<\lambda_\gamma^0<0.
13$ \\[1.mm] \hline
& & \\[-3.mm]
$p\bar p\to Z\gamma\rightarrow e^+e^-\gamma$ &
$\!\!\!\!\!\!\!-0.105<h^Z_{30}<0.105$ & $\!-0.044<h^Z_{30}<0.044$ \\[1.mm]
$\sqrt{s}=2$~TeV, $\Lambda_{FF}=1.5$~TeV &
$\!\!\!\!\!\!\!-0.0064<h^Z_{40}<0.0064$ & $\!-0.0025<h^Z_{40}<0.0025$
\\[1.mm] \hline
\end{tabular}
\end{center}
\end{table}
Future experiments at the Tevatron can measure $\Delta\kappa_\gamma$
with a precision of about 0.1~--~0.2. $\lambda_\gamma$ can be determined
to better than about 0.1 for $\int\!{\cal L}dt\ge 1$~fb$^{-1}$. The limits
for $Z\gamma V$ couplings are of order $10^{-2}~-~10^{-3}$.

\vglue 0.2cm
{\sl\noindent 3.2.3 Di-boson Production at the LHC}
\vglue 0.2cm

Since terms proportional to the non-standard $WWV$ and $Z\gamma V$
couplings in the
di-boson production amplitudes grow with energy like a power of
$\sqrt{\hat s}/m_W$, one expects~\cite{BZ1} that experiments at the LHC
($pp$ collisions at $\sqrt{s}=14$~TeV; ${\cal L}=1.7\cdot
10^{34}$~cm$^{-2}$~s$^{-1}$~\cite{LHC}) will be able to improve
significantly the limits which can be obtained at the Tevatron. To
simulate the sensitivity of $W\gamma$ and $Z\gamma$ production at the
LHC to non-standard three vector boson couplings, we use
the photon, electron and $E\llap/_T$ resolutions of the current ATLAS
detector design~\cite{ATLAS}. Similar results are obtained if
CMS~\cite{CMS} specifications are employed. Only $W\to e\nu_e$ and
$Z\to e^+e^-$ decays are studied. Acceptances are obtained using the
following set of cuts:
\begin{eqnarray}
E_T(e) > 40~{\rm GeV,} & \qquad & E\llap/_T > 25~{\rm GeV,}
\label{EQ:SIM1}\\
E_T(\gamma) > 25~{\rm GeV,} &\qquad & \Delta R(e,\gamma) > 0.7,
\label{EQ:SIM2}\\
m_T^W>50~{\rm GeV,} &\qquad & m_T(e\gamma;e\llap/_T)>90~{\rm GeV,}
\label{EQ:SIM3}
\end{eqnarray}
and
\begin{eqnarray}
m(e^+e^-)>70~{\rm GeV,} &\qquad & m(e^+e^-\gamma)>110~{\rm GeV.}
\label{EQ:SIM4}
\end{eqnarray}
Since most of the sensitivity
to anomalous couplings originates from the high $E_T$ tail, the limits
which can be obtained change only very little if the $E_T(\gamma)$
($E\llap/_T$) cut
is raised to 50~--~100~GeV (40~--~50~GeV). For the electron and photon
identification efficiencies, the values obtained in the current CDF
analysis were used. The systematic uncertainty from
the integrated luminosity, parton densities, and higher order QCD
corrections was varied between 5\% and 10\%. NLO QCD corrections are
known to be large at LHC energies, and significantly reduce the
sensitivity to anomalous couplings, unless a jet veto is
imposed~\cite{NLOWGAM}. All jets in $W\gamma$ and
$Z\gamma$ events with a transverse energy larger than 50~GeV were
therefore vetoed. This cut also helps to reduce to an acceptable level
the background from $pp\to\bar tt\gamma\to W\gamma +X$
and, together with the photon and lepton isolation cuts, the $pp\to\bar
bb\gamma$ background~\cite{MOR,BS}.

The 95\% CL limit contours from a binned likelihood fit of the photon
$E_T$ distribution for an integrated luminosity of 100~fb$^{-1}$
are shown in Fig.~9. To obtain the results shown in this figure, we have
assumed a systematic uncertainty of 5\%. Almost identical curves are
obtained if the systematic uncertainty is increased to 10\%. In contrast
to the sensitivities obtained at Tevatron energies, the limits on
$WW\gamma$ couplings found for $pp$ collisions at $\sqrt{s}=14$~TeV
depend non-negligibly on the form factor scale. The bounds on
$\Delta\kappa_\gamma^0$ ($\lambda_\gamma^0$) are about a factor~3
to~4 ($\sim 10$) better than those possible at the Tevatron with
10~fb$^{-1}$. The limits on $Z\gamma V$ couplings can be improved by a
factor $\sim 10$ ($h^V_{30}$) to $\sim 30$ ($h^V_{40}$) for
$\Lambda_{FF}=1.5$~TeV. The 95\% CL limit contours for the
$Z\gamma\gamma$ couplings are almost identical to those found for
$h_{30}^Z$ and $h_{40}^Z$ and are therefore not shown in Fig.~9b. The
limits obtained for $Z\gamma V$ couplings
depend very strongly on the value of $\Lambda_{FF}$ assumed. Increasing
the form factor scale from 1.5~TeV to 3~TeV, the limits improve by a
factor~5 to~10. The results shown in
Fig.~9 can be improved by about 20~--~40\% if $W\to\mu\nu$ and
$Z\to\mu^+\mu^-$ decays are included in the analysis.
\begin{figure}[t]
\vskip 7.5cm
\includegraphics{fig8a.ps}
\includegraphics{fig8b.ps}
{\sl\noindent Figure~9: 95\% CL sensitivity limits for a) $WW\gamma$
couplings from $W\gamma$ production and b) $ZZ\gamma$ couplings from
$Z\gamma$ production at the LHC. Results are displayed for an integrated
luminosity of 100~fb$^{-1}$ and two different form factor scales.}
\end{figure}
The limits on anomalous $Z\gamma V$ couplings could be further
strengthened if the $Z\to\bar\nu\nu$ decay can be utilized.

Using the NLO calculation of Ref.~\cite{UJH}, sensitivity limits for
the reaction $pp\to W^\pm Z\to\ell_1^\pm\nu_1\ell_2^+\ell_2^-$ were
estimated by
performing a $\chi^2$ fit to the $p_T(Z)$ distribution. No complete detector
simulation was carried out, however, a transverse momentum cut of 25~GeV and
a rapidity cut of $|\eta(\ell_{1,2})|<2.5$, $\ell_{1,2}=e,\,\mu$, were
imposed on the charged
leptons, together with a missing transverse energy cut of 50~GeV. The
relatively large $E\llap/_T$ cut was chosen to reduce backgrounds
{\elevenit e.g} from event pileup which at LHC luminosities may result
in a non-negligible amount of ``fake'' missing transverse
energy~\cite{PILE}, and from processes such as $pp\to Zb\bar
b\to\ell_1\nu_1\ell^+_2\ell_2^-+X$. The
large $E\llap/_T$ cut has only very little impact on the sensitivity
limits which can be achieved. In
addition, leptons of the same charge are required to be separated by
$\Delta R>0.4$. To reduce the effect of QCD corrections, and the
$pp\to t\bar t\to\ell_1\ell_2\ell_2 + X$~\cite{ATLAS} and
$pp\to\bar ttZ$~\cite{MOR} backgrounds on the
sensitivity limits, jets with $p_T(j)>50$~GeV and $|\eta(j)|<2.5$ are
vetoed. Particle momenta are smeared according to the resolution
expected for the ATLAS detector~\cite{ATLAS}. A 50\% normalization
uncertainty of the SM $p_T(Z)$ distribution was taken into account in
the derivation of the 95\% CL limit contours, which are shown in Fig.~10
for $\int\!{\cal L}dt=100$~fb$^{-1}$ and two choices for the
form factor scale.
\begin{figure}
\vskip 7.5cm
\includegraphics{fig9a.ps}
\includegraphics{fig9b.ps}
{\sl\noindent Figure~10: 95\% CL sensitivity limits from $W^\pm Z\to
\ell_1^\pm\nu_1\ell_2^+\ell_2^-$ at the LHC a) in the HISZ scenario and
b) if only $\Delta\kappa_Z$ and $\lambda_Z$ are allowed to deviate from
the SM.}
\end{figure}
\begin{table}
\caption{Expected 95\% CL limits on anomalous $WWV$, $V=\gamma, \, Z$,
and $ZZ\gamma$ couplings from experiments at the LHC ($pp$ collisions at
$\protect{\sqrt{s}=14}$~TeV; $\protect{\int\!{\cal
L}dt=100}$~fb$\protect{^{-1}}$). Only one of the
independent couplings is assumed to deviate from the SM at a
time. The limits obtained for
$Z\gamma\gamma$ couplings almost coincide with those found for $h_3^Z$
and $h_4^Z$.}
\begin{center}
\begin{tabular}{|ccc|}
\hline
& & \\[-3.mm]
channel & limit & limit\\[1.mm]
 & $\Lambda_{FF}=3$~TeV & $\Lambda_{FF}=10$~TeV \\[1.mm] \hline
& & \\[-3.mm]
$pp\to W^\pm\gamma\rightarrow e^\pm\nu\gamma$ &
$\!\!\!\!\!\!\!\!-0.080<\Delta\kappa_\gamma^0<0.080$ &
$\!\!-0.065<\Delta\kappa_\gamma^0<0.065$ \\[1.mm]
& $\!\!\!\!\!\!\!\!-0.0057<\lambda_\gamma^0<0.0057$ &
$\!\!-0.0032<\lambda_\gamma^0<0.0032$ \\[1.mm] \hline
& & \\[-3.mm]
$pp\to W^\pm Z\to\ell_1^\pm\nu_1\ell_2^+\ell_2^-$ & $\!\!\!\!\!\!\!\!-0.0060
<\Delta\kappa_\gamma^0<0.0097$ & $\!\!-0.0043 <
\Delta\kappa_\gamma^0<0.0086$ \\[1.mm]
$\ell_{1,2}=e,\,\mu$, HISZ scenario &
$\!\!\!\!\!\!\!\!-0.0053<\lambda_\gamma^0<0.0067$ &
$\!\!-0.0043<\lambda_\gamma^0<0.0038$ \\[1.mm] \hline
& & \\[-3.mm]
$pp\to W^\pm Z\to\ell_1^\pm\nu_1\ell_2^+\ell_2^-$ & $\!\!\!\!\!\!\!\!-0.064
<\Delta\kappa_Z^0<0.107$ & $\!\!-0.050 < \Delta\kappa_Z^0<0.078$ \\[1.mm]
$\ell_{1,2}=e,\,\mu$, $\Delta g_1^Z=0$&
$\!\!\!\!\!\!\!\!-0.0076<\lambda_Z^0<0.0075$ &
$\!\!-0.0043<\lambda_Z^0<0.0038$ \\[1.mm] \hline
& & \\[-3.mm]
channel & limit & limit\\[1.mm]
 & $\!\!\!\!\!\!\!\!\Lambda_{FF}=1.5$~TeV & $\!\!\Lambda_{FF}=3$~TeV
\\[1.mm] \hline
& & \\[-3.mm]
$pp\to Z\gamma\rightarrow e^+e^-\gamma$ &
$\!\!\!\!\!\!\!\!-0.0051<h^Z_{30}<0.0051$ & $\!\!-0.0013<h^Z_{30}<0.
0013$ \\[1.mm]
& $\!\!\!\!\!\!\!\!-9.2\cdot10^{-5}<h^Z_{40}<9.2\cdot10^{-5}$ &
$\!\!-6.8\cdot 10^{-6}<h^Z_{40}<6.8\cdot 10^{-6}$ \\[1.mm] \hline
\end{tabular}
\end{center}
\end{table}
In Fig.~10a we show 95\% CL
limits for the HISZ scenario [see Eqs.~(\ref{EQ:HISZ1})
--~(\ref{EQ:HISZ3})]. Figure~10b displays sensitivity bounds for the case
where only $\Delta\kappa_Z$ and $\lambda_Z$ are varied.

At the LHC, the $t\bar t$ production rate for top quark masses in the
range from 150~GeV to 200~GeV is about a factor~10 to~30 larger than
the $pp\to W^+W^-$ cross section~\cite{TOPLHC}. Unless the top quark
background can be reduced very efficiently, one does not expect that
$W^+W^-$ and semihadronic $WZ$ production yield limits on
anomalous $WWV$ couplings which can compete with those obtained from
$pp\to W\gamma$ and double leptonic $WZ$ decays.

Table~4 compares
the sensitivities which can be achieved in $W\gamma$, $WZ$ and $Z\gamma$
production at the LHC with 100~fb$^{-1}$. If the integrated luminosity
is reduced by a factor~10, the bounds listed in Table~4 are weakened by
about a factor~2. $\Delta\kappa_V$ and $\lambda_V$ in general can be
probed to better than 0.1 and 0.01 at the LHC, respectively.
The limits which
are obtained in the HISZ scenario for $\Delta\kappa_\gamma$ from $W^\pm
Z\to\ell_1^\pm\nu_1\ell_2^+\ell_2^-$ are ${\cal O}(10^{-2})$ and thus
much stronger than those
from $W\gamma$ production. This is due to the relation
between $\Delta g_1^Z$ and $\Delta\kappa_\gamma$ in the HISZ scenario [see
Eq.~(\ref{EQ:HISZ1})], and the fact that the terms proportional to
$\Delta g_1^Z$ in $WZ$ production grow like $\hat s/m_W^2$ with energy,
whereas the terms proportional to $\Delta\kappa_\gamma$ in $W\gamma$
production only increase like $\sqrt{\hat s}/m_W$ at most. Varying the
form factor scale from 3~TeV to 10~TeV, the limits on $WWV$ couplings
improve by about 30\%. For $\Lambda_{FF}<3$~TeV, the bounds deteriorate
rather quickly; for $\Lambda_{FF}=1$~TeV they are a factor 2~--~5 weaker
than those found for $\Lambda_{FF}=3$~TeV. As mentioned before, the
sensitivities obtained
for $Z\gamma V$ couplings depend even more strongly on the form factor
scale. A maximum scale of $\sim 10$~TeV can be probed in $W\gamma$ and
$WZ$ production, whereas scales up to 6~TeV are accessible in $Z\gamma$
production at the LHC. The limits from $W\gamma$ and $WZ$ production
listed in Table~4 are consistent with those found in Ref.~\cite{ATLAS}.

\vglue 0.2cm
{\sl\noindent 3.2.4 Amplitude Zeros and Rapidity Correlations in
$W\gamma$ and $WZ$ Production}
\vglue 0.2cm

$W\gamma$ and $WZ$ production in hadronic collisions are of
special interest due to the presence  of amplitude zeros. It is well
known that all SM helicity amplitudes of the
parton-level subprocess $q_1\bar q_2 \rightarrow W^\pm\gamma$ vanish
for~\cite{WGZERO}
\begin{eqnarray}
\cos\theta={Q_1+Q_2\over Q_1-Q_2}~,
\end{eqnarray}
where $\theta$ is the scattering angle of the $W$-boson with respect to
the quark ($q_1$) direction in the $W\gamma$ rest frame, and $Q_i$
($i=1,2$) are the quark charges in units of the proton electric charge
$e$. This zero is a consequence of the factorizability~\cite{ZHU} of
the amplitudes in gauge theories into one factor which contains the
gauge coupling dependence and another which  contains spin information.
Although the factorization holds for any four-particle Born-level
amplitude in which one or more of the four particles is a gauge-field
quantum, the amplitudes for most processes may not necessarily
develop a  kinematical zero in the physical region. The amplitude zero
in the $W^\pm \gamma$ process has been further shown to
correspond to the absence of dipole radiation by colliding particles
with the same charge-to-mass ratio~\cite{BROD}, a realization of
classical radiation interference.

Recently, it was found~\cite{WZZERO} that the SM amplitude of the process
$q_1\bar q_2 \rightarrow W^\pm Z$ also exhibits an approximate zero at high
energies. The $(\pm,\mp)$ amplitudes ${\cal M}(\pm,\mp)$ vanish for
\begin{eqnarray}
{g^{q_1}_{-} \over \hat u} + {g^{q_2}_{-} \over \hat t} = 0,
\end{eqnarray}
where $g^{q_i}_{-}$ is the coupling of the $Z$ boson to left-handed
quarks, and $\hat u$ and $\hat t$ are Mandelstam variables
in the parton center of mass frame. For $\hat s\gg m_Z^2$, the zero in
the $(\pm,\mp)$ amplitudes is located at $\cos\theta_0
=  (g^{q_1}_{-} + g^{q_2}_{-}) /  (g^{q_1}_{-} - g^{q_2}_{-})$, or
\[ \cos\theta_0 \simeq \left\{
\begin{array}{ll}
+{1\over 3}\tan^2\theta_{\rm w} \simeq +0.1
& \mbox{for $d \bar u \rightarrow W^{-} Z\>,$} \\[1.mm]
 - {1\over 3}\tan^2\theta_{\rm w}  \simeq -0.1
& \mbox{for $u \bar d \rightarrow W^{+} Z\>.$}
\end{array}
\right. \]
The existence of the zero in ${\cal M}(\pm,\mp)$ at $\cos\theta_0$ is
a direct consequence of the contributing Feynman diagrams and the
left-handed coupling of the $W$-boson to fermions.

At high energies, strong cancellations occur, and, besides ${\cal
M}(\pm,\mp)$, only the $(0,0)$ amplitude remains non-zero. The combined
effect of the zero in ${\cal M}(\pm,\mp)$ and the gauge
cancellations at high energies in the remaining helicity amplitudes
results in an approximate zero for the $q_1\bar q_2\rightarrow W^\pm Z$
differential cross section at $\cos\theta\approx\cos\theta_0$.

Non-standard $WWV$ couplings in general destroy the amplitude zeros in
$W\gamma$ and $WZ$ production. Searching for the amplitude zeros thus
provides an additional test of the gauge theory nature of the SM.

Unfortunately,
the radiation zero in $q_1\bar q_2\to W\gamma\to\ell\nu\gamma$ and the
approximate amplitude zero in $q_1\bar q_2\to
WZ\to\ell_1\nu_1\ell_2^+\ell_2^-$ are not easy to observe in the
$\cos\theta$ distribution in $pp$ or
$p\bar p$ collider experiments. Structure function effects transform
the zero in the $W\gamma$ case into a dip in the
$\cos\theta$ distribution. The approximate zero in $WZ$ production is
only slightly affected by structure function effects. Higher order QCD
corrections~\cite{JACK1} and finite $W$ width effects~\cite{JACK2} tend
to fill in the
dip. In $W\gamma$ production photon radiation from the final state
lepton line also diminishes the significance of the effect.

The main complication in the extraction of the $\cos\theta$
distribution, however, originates from the finite resolution of the
detector and
ambiguities in reconstructing the parton center of mass frame. The
ambiguities are associated with the
nonobservation of the neutrino arising from $W$ decay. Identifying the
missing transverse momentum with the transverse momentum of the neutrino
of a given $W\gamma$ or $WZ$ event, the unobservable longitudinal neutrino
momentum, $p_L(\nu)$, and thus the parton center of
mass frame, can be reconstructed by imposing the constraint that the
neutrino and charged lepton four momenta combine to form the $W$ rest
mass~\cite{GUN}. The resulting quadratic equation, in general, has
two solutions. In the approximation of a zero $W$ decay width, one of
the two solutions coincides with the true $p_L(\nu)$. On an event to
event basis, however, it is impossible to tell which of the two
solutions is the correct one. This ambiguity
considerably smears out the dip caused by the amplitude zeros.

Instead of trying to reconstruct the parton center of mass frame and
measure the $\cos\theta$ or the equivalent rapidity distribution in the
center of mass frame, one
can study rapidity correlations between the observable final state
particles in the laboratory frame~\cite{BEL}. Knowledge of the neutrino
longitudinal momentum is not required in determining the rapidity
correlations.
Event mis-reconstruction problems originating from the two possible
solutions for $p_L(\nu)$ are thus automatically avoided.

In $2\to 2$ reactions differences of rapidities are invariant
under boosts. One therefore expects that the rapidity difference
distributions $d\sigma/d\Delta y(V,W)$, $V=\gamma,\, Z$, where $\Delta
y(V,W)=y(V)-y(W)$ and $y(W)$, $y(V)$ are the rapidities in the
laboratory frame, exhibit a dip signaling the SM amplitude
zeros~\cite{BEL}. In
$W^\pm\gamma$ production, the dominant $W$ helicity is $\lambda_W = \pm
1$~\cite{BIL}, implying that
the charged lepton, $\ell=e,\,\mu$, from $W\to\ell\nu$ tends to be emitted
in the direction of the parent
$W$, and thus reflects most of its kinematic properties. As a result,
the dip signaling the SM radiation zero should manifest itself in the
$\Delta y(\gamma,\ell)=y(\gamma)-y(\ell)$ distribution.

\begin{figure}[t]
\vskip 10.cm
\includegraphics{fig10.ps}
{\sl\noindent Figure~11: Rapidity difference distributions in the SM at the
Tevatron. a) The photon lepton rapidity difference spectrum in $p\bar
p\to\ell^+p\llap/_T\gamma$. b) The $y(Z)-y(\ell_1^+)$ and $y(\ell_2^+)
-y(\ell_1^+)$ distributions in $p\bar p\to W^+Z$. }
\end{figure}
The SM $\Delta y(\gamma,\ell)$ differential cross section for $p\bar
p\to\ell^+p\llap/_T\gamma$ at the Tevatron is shown in Fig.~11a. To
simulate detector response, transverse momentum cuts of $p_T(\gamma)
>5$~GeV, $p_T(\ell)>20$~GeV and $p\llap/_T>20$~GeV, rapidity cuts of
$|y(\gamma)|<3$ and $|y(\ell)|<3.5$, a cluster transverse mass cut of
$m_T(\ell\gamma;p\llap/_T)>90$~GeV and a lepton photon separation cut of
$\Delta R(\gamma,\ell)>0.7$ have been imposed. The SM radiation zero is
seen to lead to a strong dip in the $\Delta y(\gamma,\ell)$ distribution
at $\Delta y(\gamma,\ell)\approx -0.3$. Next-to-leading QCD corrections
do not seriously affect the significance of the dip. However, a sufficient
rapidity coverage is essential to observe the radiation zero in the
$\Delta y(\gamma,\ell)$ distribution~\cite{BEL}.

In contrast to the situation in $W\gamma$ production, none of the $W$
helicities dominates
in $WZ$ production~\cite{BIL}. The charged lepton originating from the $W$
decay, $W\to\ell_1\nu_1$, thus only partly reflects the kinematical
properties of the parent $W$ boson. As a result, a significant part of
the correlation present in the $y(Z)-y(W)$ spectrum~\cite{FNR} is lost,
and only a slight dip survives in the $y(Z)-y(\ell_1)$ distribution, which is
shown for the $W^+Z$ case in Fig.~11b. The dip in the SM $y(Z)-y(\ell_1)$
distribution will thus be more difficult to observe experimentally than
that in the $y(\gamma)-y(\ell)$ distribution in $W\gamma$ production.
Next-to-leading order QCD
corrections have only little impact on the shape of the $y(Z)-y(\ell_1)$
distribution~\cite{UJH}. The cuts used in Fig.~11b are the same as
those in Fig.~3a except for the lepton rapidity cut which has been
replaced by $|y(\ell_{1,2})|<2.5$.

Although the $Z$ boson rapidity,
$y(Z)$, can readily be reconstructed from the four momenta of the
lepton pair $\ell_2^+\ell_2^-$ originating from the $Z$ decay, it would
be easier experimentally to directly study
the rapidity correlations between the charged leptons
originating from the $Z\to\ell_2^+\ell_2^-$ and $W\to\ell_1\nu_1$
decays. The dotted line in Fig.~11b shows the $y(\ell_2^+)-y(\ell_1^+)$
distribution for $W^+Z$ production at the Tevatron. The $y(\ell_2^-)
-y(\ell_1^+)$ spectrum almost coincides with the
$y(\ell_2^+)-y(\ell_1^+)$ distribution. Since also none of the $Z$ boson
helicities dominates~\cite{BIL} in $q_1\bar q_2\to WZ$, the
rapidities of the leptons from $W$ and $Z$ decays are almost completely
uncorrelated, and essentially no trace of the dip signaling the approximate
amplitude zero is left in the $y(\ell_2^+)-y(\ell_1^+)$ distribution.

In $pp$ collisions, the dip signaling the amplitude zeros is shifted to
$\Delta y=0$. Because of the large $qg$ luminosity, the inclusive QCD
corrections are very large for $W\gamma$ and $WZ$
production~\cite{UJH,NLOWGAM}. At the
LHC, they enhance the cross section by a factor 2~--~3. The rapidity
difference distributions for $W^+\gamma$ and $W^+Z$ production in the SM
for $pp$ collisions at $\sqrt{s}=14$~TeV are shown in Fig.~12. Here we have
imposed the following lepton and photon detection cuts:
\begin{eqnarray}
p_T(\gamma) > 100~{\rm GeV}, & \qquad & |\eta(\gamma)|<2.5, \\
p_T(\ell)>25~{\rm GeV}, &\qquad &|\eta(\ell)|<3, \\
p\llap/_T>50~{\rm GeV}, &\qquad &\Delta R(\gamma, \ell)>0.7,
\end{eqnarray}
together with a $\Delta R(\ell,\ell)>0.4$ requirement on leptons of the
same charge in $WZ$ production.
\begin{figure}[t]
\vskip 10.cm
\includegraphics{fig11.ps}
{\sl\noindent Figure~12: Rapidity difference distributions in the SM at
the LHC. a) The photon lepton rapidity difference spectrum in $p
p\to\ell^+p\llap/_T\gamma$. b) The $y(Z)-y(\ell_1^+)$ distribution
in $pp\to W^+Z$. }
\end{figure}
The inclusive NLO QCD corrections are seen to considerably obscure
the amplitude zeros. The bulk of the corrections at LHC energies
originates from quark gluon fusion and the kinematical region where
{\elevenit e.g.} the
photon or $Z$ boson is produced at large $p_T$ and recoils against a
quark, which radiates a soft $W$ boson which is almost collinear to the
quark. Events which originate from this phase space region usually
contain a high $p_T$ jet. A jet veto therefore helps to reduce the QCD
corrections, as demonstrated by the dotted lines in Fig.~12. Here a jet
is defined as a quark or gluon with $p_T(j)>50$~GeV and $|\eta(j)|<3$.
Nevertheless, the remaining QCD corrections
still substantially reduce the visibility of the radiation zero in
$W\gamma$ production at the LHC. In $pp\to WZ$, the difference in
significance of the dip between the LO and the NLO 0-jet $\Delta y(Z,
\ell_1)$ distribution is quite small.

Given a sufficiently large integrated luminosity, experiments at the
Tevatron studying lepton photon rapidity correlations offer a much
better chance to observe the SM radiation zero in $W\gamma$ production
than experiments at the LHC. Searching for the approximate amplitude
zero in $WZ$ production will be difficult at the Tevatron as well as the
LHC.

Indirectly, the radiation zero can also be observed in the $Z\gamma$ to
$W\gamma$ cross section ratio~\cite{BEO}. Many theoretical and
experimental uncertainties at least partially cancel in the cross section
ratio. On the other hand, in searching for the effects of the SM
radiation zero in the $Z\gamma$ to $W\gamma$ cross section ratio, one
has to assume that the SM is valid for $Z\gamma$ production. Similarly,
the $ZZ$ to $WZ$ cross section ratio reflects the approximate amplitude
zero in $WZ$ production, whereas the ratio of $WZ$ to $W\gamma$ cross
sections measures the relative strength of the zeros in $WZ$ and
$W\gamma$ production~\cite{UJH}.

\vglue 0.2cm
{\elevenit\noindent 3.3 Probing $WWV$ and $Z\gamma V$ Couplings in
$e^+e^-$ Collider Experiments}
\vglue 0.2cm
{\sl\noindent 3.3.1 Single Photon Production at LEP}
\vglue 0.1cm

In $e^+e^-$ collisions at center of mass energies near the $Z$ boson
mass, anomalous $Z\gamma V$ couplings would affect the production of
$f\bar f\gamma$ final states. At LEP energies, the production of single
photons is the process which is most sensitive to anomalous $ZZ\gamma$
couplings, due to the large branching ratio for $Z\to\nu\bar\nu$ decays
and the absence of background from final state radiation or final state
$\pi^0$'s misidentified as photons. In order to probe $Z\gamma\gamma$
couplings one has to study $\ell^+\ell^-\gamma$ or $jj\gamma$ final
states.

The L3~Collaboration has searched for anomalous $ZZ\gamma$ couplings in
single photon events in the data collected in 1991~--~93~\cite{L3}.
Non-standard $ZZ\gamma$ couplings mostly affect the production of
energetic single photon events whereas the photon energy spectrum in the SM
process $e^+e^-\to\bar\nu\nu\gamma$ is peaked at low energies. Therefore,
a cluster
in the BGO electromagnetic calorimeter with energy greater than half the
beam energy was required. In order to further reduce the SM
contribution and to eliminate the background from QED events in which
all final state particles except the photon escape undetected down the
beampipe or into a detector crack, it was required that the polar angle of
the most energetic cluster lies between 20 and 160 degrees (excluding
the ranges between 34.5 and 44.5, and 135.5 and 145 degrees due to gaps
between the forward and barrel BGO calorimeters). To suppress the
background from cosmic events, the transverse shape of the BGO cluster
was required to be consistent with a photon originating from the
interaction point. Apart from the energetic BGO cluster, all other
activity in the detector had to be consistent with noise. In
terms of equivalent integrated luminosity at the peak of the $Z$
resonance, the data sample corresponds to 50.8~pb$^{-1}$.
One event was selected. The number of events expected in the SM is~1.2.

Since the level of energetic single photon production is consistent with
what is expected in the SM, upper limits on the $ZZ\gamma$ couplings can
be derived. To extract limits, a modified version of the event
generator of Ref.~\cite{UBELB2} was used. Events were generated for
various combinations of $ZZ\gamma$ couplings, and passed through the
detector simulation and analysis procedure. Figure~13 shows the 95\% CL
upper limits on $h_{30}^Z$ and $h_{40}^Z$ for a form factor scale of
$\Lambda_{FF}=500$~GeV. Also shown are the current limits from D\O\ and
CDF. Table~5 summarizes the numerical values, if only one of the
couplings deviates from the SM at a time.
\begin{figure}
\vskip 7.5cm
\includegraphics{fig12.ps}
{\sl\noindent Figure~13: Present limits on anomalous $ZZ\gamma$
couplings from $Z\to\bar\nu\nu\gamma$, and from $Z\gamma$ production at
the Tevatron.}
\end{figure}
\begin{table}
\caption{95\% CL limits on anomalous $Z\gamma V$, $V=\gamma,
\, Z$, couplings from L3, CDF and D\O. Only one of the independent
couplings is allowed to deviate from the SM at a time. The form
factor scale is chosen to be $\Lambda_{FF}=500$~GeV.}
\begin{center}
\begin{tabular}{|ccc|}
\hline
& & \\[-3.mm]
experiment & channel & limit \\[1.mm] \hline
& & \\[-3.mm]
L3 & $e^+e^-\to Z\to\bar\nu\nu\gamma$ & $-0.85 < h^Z_{30} < 0.85$ \\[1.mm]
 & & $ -2.32 <h^Z_{40} < 2.32$ \\[1.mm]\hline
& & \\[-3.mm]
CDF & $p\bar p\rightarrow Z\gamma\rightarrow\ell^+\ell^-\gamma$ &
$-3.0<h^Z_{30}<2.9$ \\[1.mm]
 & $\ell=e,\,\mu$ & $-0.7<h^Z_{40}<0.7$ \\[1.mm] \hline
& & \\[-3.mm]
D\O\ & $p\bar p\rightarrow Z\gamma\rightarrow\ell^+\ell^-\gamma$ &
$-1.9<h^Z_{30}<1.8$ \\[1.mm]
 & $\ell=e,\,\mu$ & $-0.5<h^Z_{40}<0.5$ \\[1.mm] \hline
\end{tabular}
\end{center}
\end{table}
The limits obtained from $Z\to\bar\nu\nu\gamma$ on $h^Z_{30}$ are
significantly better than those found from $Z\gamma$ production at the
Tevatron. On the other hand, because of the larger center of mass
energy and the
strong increase of the terms proportional to $h_4^Z$ in the helicity
amplitudes, hadron collider experiments give much better bounds on
$h_{40}^Z$ than single photon production at LEP. LEP and Tevatron
experiments thus yield complementary information on $ZZ\gamma$
couplings. LEP will discontinue to run on the $Z$ peak in 1996.
Final integrated statistics are expected to increase by perhaps a
factor~3 over that used in the current analysis. Consequently,
the present limits on $h_{30}^Z$ and $h_{40}^Z$ from
$Z\to\bar\nu\nu\gamma$ are expected to improve by not more than about
a factor~2 in the future. In contrast to
the limits obtained from hadron collider experiments, the sensitivity
bounds derived from $Z\to\bar\nu\nu\gamma$ only marginally depend on the
form factor scale.

An analysis of $\ell^+\ell^-\gamma$ final states is in progress.

\vglue 0.2cm
{\sl\noindent 3.3.2 $W^+W^-$ and $Z\gamma$ Production at LEP~II}
\vglue 0.2cm

$W$ pair production and $Z\gamma$ production at LEP~II ($\sqrt{s}=176 -
190$~GeV) offer ideal
possibilities to probe $WWV$~\cite{HPZH,Sek,SCH} and $Z\gamma V$~\cite{CYZ}
couplings. In contrast to
$pp,\, p\bar p\to W^+W^-\to\ell\nu jj$, the reaction $e^+e^-\to
W^+W^-\to\ell\nu jj$ is not plagued by large backgrounds. Furthermore, the
reconstruction of the leptonically decaying $W$ boson is easier than in
hadronic collisions, where the longitudinal momentum of the neutrino can
be reconstructed only with a twofold ambiguity. At hadron colliders,
limits on non-standard couplings are derived from distributions such as
the transverse momentum distribution of one of the vector bosons which
make use of the high energy behaviour of the anomalous contributions to
the helicity amplitudes. At LEP~II, on the other hand, angular
distributions are more useful. Different anomalous couplings contribute
to different helicity amplitudes and therefore affect the angular
distributions in a characteristic way (see Ref.~\cite{HPZH}).

In $W^+W^-$ production, 5~angles are available from each event. These
are the $W$ production angle, $\Theta_W$, and the angles of the
$W^\pm\to f\bar f'$ decay products in the $W^\pm$ rest frames,
$\theta_\pm$ and $\phi_\pm$. In the extraction of these angles,
two problems have to be faced: First, the imperfect detection of $W$
decay products gives rise to uncertainties in the reconstructed
directions of the $W$'s and their decay products; second, in the case of
hadronic $W$ decays, the absence of a readily recognizable quark tag
implies that the $W$ decay angles can only be determined with a two-fold
ambiguity from the data, resulting in symmetrized angular
distributions. Complete information for a $W^+W^-$ event is only
available if it is possible to distinguish the $W^+$ and $W^-$ direction.

Of the three final states available in $W$ pair production,
$\ell_1\nu_1\ell_2\nu_2$, $\ell\nu jj$, $\ell=e,\,\mu$, and $jjjj$, we
have only
studied the $\ell\nu jj$ channel. The purely leptonic channel is plagued
by a small branching ratio ($\approx 4.7\%$) and by reconstruction
problems due to the presence of two neutrinos. In the $jjjj$ final state
it is difficult to discriminate the $W^+$ and $W^-$ decay products. Due
to the resulting ambiguities in $\Theta_W$ and the $W^\pm$ decay angles,
the sensitivity bounds which can be achieved from the 4-jet final state
are a factor 1.5 --~2 weaker than those found from analyzing the
$\ell\nu jj$ state~\cite{Sek}. In
the $\ell\nu jj$ channel, on the other hand, the identification of the
charged lepton allows the $W^+$ and $W^-$ decays to be distinguished
unambiguously.

Events in the $\ell\nu jj$ channel were selected from simulated Monte
Carlo data at $\sqrt{s}=176$~GeV and $\sqrt{s}=190$~GeV using the event
generator of Ref.~\cite{HPZH}. Initial state radiation and detector
smearing, using the L3 specifications are taken into account in the
simulations. The following cuts were imposed:
\begin{itemize}
\item Number of calorimetric clusters $>16$. This requirement eliminates
almost all $WW\to\ell_1\nu_1\ell_2\nu_2$, $\ell_{1,2}=e,\,\mu$ events.
It also helps to suppress the $WW\to\tau\nu_\tau\ell\nu$, $\ell=e,\,
\mu$ and $WW\to\tau\nu_\tau\tau\nu_\tau$ channels where at least one of the
$\tau$ leptons decays hadronically. Furthermore it provides some
rejection of $e^+e^-\to\gamma\gamma$ and $e^+e^-\to\tau^+\tau^-(\gamma)$
events.

\item A visible energy $E_{vis}>80$~GeV. This cut mainly reduces the
background from $e^+e^-\to\gamma\gamma$ and
$e^+e^-\to\tau^+\tau^-(\gamma)$, removes signal events which are poorly
reconstructed, and further suppresses $WW\to\tau\nu_\tau\ell\nu$, $\ell=e,\,
\mu$ and $WW\to\tau\nu_\tau\tau\nu_\tau$ events where at least one of the
$\tau$ leptons decays hadronically.

\item $E\llap/_T/E_{vis}>0.1$. It reduces the $WW\to jjjj$ and
$Z/\gamma^*(\gamma)\to jj(\gamma)$ backgrounds. Here, ``$(\gamma)$''
denotes a photon from initial state radiation.

\item The momentum of the most energetic lepton, positively identified
as an electron or muon, is $p_{max}>20$~GeV. This cut provides most of
the suppression of the $WW\to\tau\nu_\tau jj$ and $jjjj$, and
$Z/\gamma^*\to jj(\gamma)$ backgrounds.

\item $65~{\rm GeV}<m(\ell\nu)<125$~GeV. The neutrino momentum was
calculated from momentum balance in the event. This requirement mostly
suppresses the $WW\to\tau\nu_\tau jj$ background.
\end{itemize}
With these cuts, the selection efficiency is about 70\%, and the ratio
of signal to background is approximately~20.

Sensitivities to the $WWV$ couplings are calculated for the HISZ
scenario [see Eqs.~(\ref{EQ:HISZ1}) --~(\ref{EQ:HISZ3})] from the
results of a binned maximum log likelihood fit to event distributions,
assuming an integrated luminosity of 500~pb$^{-1}$ which corresponds to
several years of running.
Figure~14a shows the 95\% CL limit contours obtained at 176~GeV and
190~GeV from a fit to the
$\cos\Theta_W$, $\cos\theta_\ell$, $\phi_\ell$, $\cos\theta_j$ and
$\phi_j$ distributions, where the (down type) jet $j$ was chosen at
random from the jet pair, {\elevenit i.e.} it was assumed that quarks
cannot be tagged. Close to the $W$ pair threshold, the gauge theory
cancellations are not fully operative and the sensitivity to anomalous
$WWV$ is limited. If the LEP~II center of mass energy can be
increased to 190~GeV, the sensitivity bounds improve by about a
factor~1.5. The limits on $\Delta\kappa_\gamma^0$ and
$\lambda_\gamma^0$ for $\sqrt{s}=176$~GeV and 190~GeV in the HISZ
scenario are summarized in Table~6 for the case where only one of the two
couplings deviates from the SM at a time.
\begin{figure}
\vskip 7.5cm
\includegraphics{fig13a.ps}
\includegraphics{fig13b.ps}
{\sl\noindent Figure~14: 95\% CL sensitivity limits from $e^+e^-\to
W^+W^-\to\ell\nu jj$ at LEP~II for an integrated luminosity of
500~pb$^{-1}$. a) Limit contours for $\sqrt{s}=176$~GeV
and 190~GeV from fitting all five angular distributions, assuming no
quark tagging. b) Contours obtained assuming that no information about
the hadronically decaying $W$ is used (dashed line), using all five
angles assuming no quark tagging (solid line), and contours found for
the hypothetical situation that all five angles are used and quarks are
tagged with 100\% efficiency (dotted line).}
\end{figure}
\begin{table}
\caption{Expected 95\% CL limits on anomalous $WWV$, $V=\gamma, \, Z$,
couplings from experiments at LEP~II in the HISZ scenario [see
Eqs.~(\protect{\ref{EQ:HISZ1}}) --~(\protect{\ref{EQ:HISZ3}})] for two
center of mass
energies. The integrated luminosity assumed is $\protect{\int\!{\cal
L}dt=500}$~pb$\protect{^{-1}}$. Only one of the
independent couplings is assumed to deviate from the SM at a
time. The limits are obtained from
a binned log likelihood fit to all five angles, assuming no quark
tagging.}
\begin{center}
\begin{tabular}{|ccc|}
\hline
& & \\[-3.mm]
dependent couplings & limit & limit\\[1.mm]
 & $\sqrt{s}=176$~GeV & $\sqrt{s}=190$~GeV  \\[1.mm] \hline
& & \\[-3.mm]
Eqs.~(\ref{EQ:HISZ1}) and~(\ref{EQ:HISZ2}) &
$-0.19<\Delta\kappa_\gamma^0<0.21$ & $-0.13<\Delta\kappa_\gamma^0<0.14$
 \\[2.mm]
$\lambda_\gamma=\lambda_Z$ & $-0.18<\lambda_\gamma^0<0.19$ &
$-0.13<\lambda_\gamma^0<0.14$ \\[1.mm] \hline
\end{tabular}
\end{center}
\end{table}
Note that the limits on $\Delta\kappa_\gamma^0$ and
$\lambda_\gamma^0$ at LEP~II are quite strongly correlated, in contrast
to those obtained from $W\gamma$ and $WW$, $WZ$ production production in
hadronic collisions. The much reduced correlations at hadron colliders
are due to the high Tevatron and LHC center of mass energies, and the
different high energy behavior of terms proportional to
$\Delta\kappa_\gamma$ and $\lambda_\gamma$ in the helicity
amplitudes. Figure~14b shows limit contours at
$\sqrt{s}=176$~GeV for binning events in $\cos\Theta_W$,
$\cos\theta_\ell$, and $\phi_\ell$ only (dashed line), all five angles
assuming that quarks cannot be tagged (solid line), and for the
hypothetical case where the quarks of the hadronically decaying $W$
boson are always tagged correctly (dotted line). The dotted line thus
corresponds to the ultimate theoretical precision with which the
anomalous couplings could be determined. Whereas the information
obtained from the hadronically decaying $W$ does not affect the limits
if only one of the two couplings is varied at a time, it reduces the
correlations between $\Delta\kappa_\gamma^0$ and $\lambda_\gamma^0$ by
approximately a factor~1.5. Due to the relatively low center of mass
energy, the limits which can be achieved at LEP~II are very insensitive
to the form factor scale and power assumed.

The contributions from $Z$
and photon exchange in $e^+e^-\to W^+W^-$ tend to cancel. Therefore, if
the $WW\gamma$ or $WWZ$ couplings only are allowed to deviate from the
SM, somewhat more stringent limits are obtained than in the HISZ
scenario used in our simulations.

Single photon production~\cite{GILLES} at LEP~II yields sensitivity
limits on
the $WWV$ couplings which are substantially weaker than those derived
from $W$ pair production. The limits estimated from single $W$
production, on the other hand, are comparable to those obtained from
$e^+e^-\to W^+W^-$~\cite{PAPA}.

$Z\gamma V$ couplings can be probed in $Z\gamma$ production at LEP~II.
To illustrate the sensitivities which might be expected, 95\% CL limit
contours for the $ZZ\gamma$ and $Z\gamma\gamma$ couplings were derived
from $e^+e^-\to Z\gamma\to\bar\nu\nu\gamma$ and $e^+e^-\to
Z\gamma\to\mu^+\mu^-\gamma$, respectively. For both processes a photon
energy $E_\gamma>60$~GeV, and $|\cos\theta_\gamma|<0.8$ was required.
For single photon production, the cut on the photon energy significantly
suppresses~\cite{GILLES} the contribution from $t$-channel $W$ exchange
to the $\bar\nu_e\nu_e\gamma$ final state, which is not included in the
calculation used. The muon scattering angle, $\theta_\mu$, in
$e^+e^-\to\mu^+\mu^-\gamma$ was required to satisfy $|\cos\theta_\mu|<0.
927$ which corresponds to the L3 angular coverage for muons at LEP~II.
Muons are also required to have $p_T(\mu)>10$~GeV and to be well
isolated from the photon; $\Delta R(\mu,\gamma)>0.35$. In addition, a
cut on the di-muon mass of $m(\mu\mu)>10$~GeV is imposed. A simplified
model of the L3 detector is used to simulate detector effects.

Sensitivity bounds are calculated from a fit to the total cross section
within cuts. The resulting 95\% limit
contours for a center of mass energy of 180~GeV, $\Lambda_{FF}=1$~TeV,
and an integrated luminosity of 500~pb$^{-1}$ are shown in Fig.~15.
Since the LEP~II center of mass energy will be significantly above the
$Z\gamma$ threshold, the bounds derived on anomalous $Z\gamma V$
couplings vary only little within the expected range of center of mass
energies expected ($\sqrt{s}=176$~GeV --~190~GeV), in contrast to the
situation encountered for $W$ pair production.
The limits on $h_{30}^V$ and $h_{40}^V$
are summarized in Table~7 for the case where only one of the two
couplings deviates from the SM at a time. For the $ZZ\gamma$ couplings,
we also include the present L3 limits from $Z\to\bar\nu\nu\gamma$ for
comparison.
\begin{figure}
\vskip 7.5cm
\includegraphics{fig14a.ps}
\includegraphics{fig14b.ps}
{\sl\noindent Figure~15: 95\% CL sensitivity limits from $e^+e^-\to
Z\gamma$ at LEP~II for an integrated luminosity of
500~pb$^{-1}$. a) Limit contours for $ZZ\gamma$ couplings from single
photon production. b) Sensitivity limits for $Z\gamma\gamma$ couplings
from $e^+e^-\to\mu^+\mu^-\gamma$.}
\end{figure}
\begin{table}
\caption{Expected 95\% CL limits on anomalous $Z\gamma V$ $V=\gamma, \, Z$,
couplings from experiments at LEP~II for $\protect{\sqrt{s}=180}$~GeV.
The integrated luminosity assumed is $\protect{\int\!{\cal
L}dt=500}$~pb$\protect{^{-1}}$. Only one of the two
couplings is assumed to deviate from the SM at a
time. For comparison, we have also included the limits on
$\protect{h^Z_{30}}$ and $\protect{h^Z_{40}}$ from
$\protect{Z\to\bar\nu\nu\gamma}$ at LEP~\protect{\cite{L3}}. The form
factor scale chosen is $\protect{\Lambda_{FF}=1}$~TeV.}
\begin{center}
\begin{tabular}{|ccc|}
\hline
& & \\[-3.mm]
reactions & \multicolumn{2}{c|}{limits} \\[1.mm] \hline
& & \\[-3.mm]
$e^+e^-\to Z\to\bar\nu\nu\gamma$ &
$-0.79<h^Z_{30}<0.79$ & $-2.08<h^Z_{40}<2.08$  \\[2.mm]
$e^+e^-\to Z\gamma\to\bar\nu\nu\gamma$ & $-0.50<h^Z_{30}<0.50$ &
$-0.45<h^Z_{40}<0.45$  \\[2.mm]
$e^+e^-\to Z\gamma\to\mu^+\mu^-\gamma$ & $-0.55<h^\gamma_{30}<0.55$ &
$-0.48<h^\gamma_{40}<0.48$  \\[1.mm]\hline
\end{tabular}
\end{center}
\end{table}
Due to the higher LEP~II center of mass energy the present limits on
$ZZ\gamma$ couplings from $Z\to\bar\nu\nu\gamma$ improve by a factor~1.6
($h^Z_3$) to 4.6 ($h^Z_4$). The improvement is more pronounced for
$h^Z_4$, due to the stronger growth with energy of the terms
proportional to $h^Z_4$ in the helicity amplitudes. The limits on
$h^V_{30}$ and $h^V_{40}$ which can be achieved are quite similar at
LEP~II. The sensitivity bounds on $Z\gamma\gamma$ couplings are about
10\% weaker than those found for $ZZ\gamma$ couplings. However, they
are expected to significantly improve, if the
angular distributions of the final state particles are analyzed instead
of the total cross section.

\vglue 0.2cm
{\sl\noindent 3.3.3 $W^+W^-$ Production at the Next Linear Collider}
\vglue 0.2cm

Since the LEP~II center of mass energy is only slightly above the $W$
pair threshold, the SM gauge cancellations are not fully operative, and the
sensitivity to anomalous gauge boson couplings is limited. Much better
limits on $WWV$ and  $Z\gamma V$ couplings will be possible at an $e^+e^-$
collider operating in the several hundred GeV range or above. Such a
machine will presumably be a linear collider. Current design studies for
such a ``Next Linear Collider'' (NLC) foresee an initial stage with a
center of mass energy of 500~GeV and a luminosity of $8\cdot
10^{33}~{\rm cm}^{-2}~{\rm s}^{-1}$. In a second stage, the energy is
increased to $\sqrt{s}=1.5$~TeV, with a luminosity of $1.9\cdot
10^{34}~{\rm cm}^{-2}~{\rm s}^{-1}$~\cite{Palmer}.

As we have mentioned in Section~3.1, such a linear collider could also
be operated as a $e\gamma$, $\gamma\gamma$ and $e^-e^-$ collider, and a
variety of processes can be used to constrain the vector boson
self-interactions at the NLC. Since the limits obtained from $W$ pair
production in the $e^+e^-$ mode~\cite{NLCWW} are comparable or better
than those obtained from other processes, we restrict ourselves to the
process $e^+e^-\to W^+W^-$ in the following.

The extraction of limits on the $WWV$ couplings at the
NLC~\cite{NLCWW,Bark} follows the
same strategy employed at LEP~II. Again, only the $\ell\nu jj$ final
state is analyzed. All five angles are used in the maximum likelihood
fits. Two cuts are imposed. First, we require
$|\cos\Theta_W|<0.8$. This ensures that the event is well within
the detector volume. The second cut forces the $W^+W^-$ invariant mass
to be within a few GeV of the nominal $e^+e^-$ center of mass energy,
and ensures that the $W^+$ and $W^-$ invariant masses each are within a
few GeV of the $W$ pole mass, $m_W$. In order to impose the second cut, we
reconstruct the mass of the leptonically decaying $W$ ($m_{W_1}$), and the
mass of the hadronically decaying $W$ ($m_{W_2}$). $m_{W_2}$ is
reconstructed by imposing four energy momentum constraints and solving
for the momentum vector of the neutrino from the leptonically decaying
$W$, and $m_{W_2}$. $m_{W_1}$ is then given by
\begin{eqnarray}
m_{W_1} & = & \left [(E_\ell+E_\nu)^2 - (\hbox{\bf p}_\ell + \hbox{\bf
p}_\nu)^2\right ]^{1/2}~.
\end{eqnarray}
We then require
\begin{eqnarray}
\chi^2<2,
\end{eqnarray}
where $\chi^2$ is defined by
\begin{eqnarray}
\chi^2={(m_{W_1}-m_W)^2\over\Gamma_W^2} + {(m_{W_2}-m_W)^2\over\Gamma_W^2}~.
\end{eqnarray}
\begin{figure}
\vskip 7.5cm
\includegraphics{fig15.ps}
{\sl\noindent Figure~16: The 95\% CL limit contours for
$\Delta\kappa_\gamma$ and $\lambda_\gamma$ from $e^+e^-\to W^+W^-$ at
$\sqrt{s}=500$~GeV with 80~fb$^{-1}$ (solid line), and at
$\sqrt{s}=1.5$~TeV with 190~fb$^{-1}$ (dashed line) for the HISZ
scenario [see Eqs.~(\ref{EQ:HISZ1}) --~(\ref{EQ:HISZ3})].}
\end{figure}
\begin{table}
\caption{Expected 95\% CL limits on anomalous $WWV$, $V=\gamma, \, Z$,
couplings from experiments at the NLC in the HISZ scenario [see
Eqs.~(\protect{\ref{EQ:HISZ1}}) --~(\protect{\ref{EQ:HISZ3}})] for two
center of mass energies and integrated luminosities. Only one of the
independent couplings is assumed to deviate from the SM at a
time. The limits are obtained from a log likelihood fit to all five
angles, assuming no quark tagging. }
\begin{center}
\begin{tabular}{|ccc|}
\hline
& & \\[-3.mm]
dependent couplings & limit & limit\\[1.mm]
 & $\sqrt{s}=500$~GeV & $\sqrt{s}=1. 5$~TeV \\[1.mm]
 & $\int\!{\cal L}dt=80$~fb$^{-1}$ & $\int\!{\cal L}dt=190$~fb$^{-1}$
 \\[1.mm] \hline
& & \\[-3.mm]
Eqs.~(\ref{EQ:HISZ1}) and~(\ref{EQ:HISZ2}) &
$-0.0024<\Delta\kappa_\gamma<0.0024$ & $-5.2\cdot 10^{-4}<
\Delta\kappa_\gamma<5.2\cdot 10^{-4}$  \\[2.mm]
$\lambda_\gamma=\lambda_Z$ & $-0.0018<\lambda_\gamma<0.0018$ &
$-3.8\cdot 10^{-4}<\lambda_\gamma<3.8\cdot 10^{-4}$ \\[1.mm] \hline
\end{tabular}
\end{center}
\end{table}
Figure~16 shows the 95\% CL contours for $\Delta\kappa_\gamma$ and
$\lambda_\gamma$ at $\sqrt{s}=500$~GeV with 80~fb$^{-1}$, and at
$\sqrt{s}=1.5$~TeV with 190~fb$^{-1}$ for the HISZ scenario [see
Eqs.~(\ref{EQ:HISZ1}) --~(\ref{EQ:HISZ3})]. The limits for the case that
only one of the two independent couplings deviates from the SM are
summarized in Table~8. Depending on the energy and integrated luminosity
of the NLC, the LEP~II limits could be improved by two to three orders
of magnitude. No form factor effects are taken into account
in the bounds listed. However, due to the fixed center of mass energy,
these effects can
easily be incorporated. They result in a simple rescaling of the limits
quoted.

The sensitivities for $Z\gamma V$ couplings are expected to be of ${\cal
O}(10^{-3})$ at the NLC~\cite{boud}.

\vglue 0.3cm
{\elevenbf\noindent 4. Conclusions}
\vglue 0.2cm

In this report, we have discussed the direct measurement of $WWV$ and
$Z\gamma V$ couplings in present and future collider experiments. These
couplings are defined through a phenomenological effective Lagrangian
[see Eqs.~(\ref{LeffWWV}) and~(\ref{EQ:ZgammaV})], analogously to the
general vector and axial vector couplings, $g_V$ and $g_A$, for the
coupling of gauge bosons to fermions. The major goal of such
experiments will be the confirmation of the SM
predictions. We have also reviewed our current theoretical understanding
of anomalous gauge boson self-interactions. If the energy scale of the
new physics responsible for the non-standard gauge boson couplings is $\sim
1$~TeV, these anomalous couplings are expected to be no larger than
${\cal O}(10^{-2})$.

Rigorously speaking, the three gauge boson vertices are unconstrained
by current electroweak precision experiments. Such experiments only
lead to bounds on the anomalous
couplings if one {\elevenit assumes} that cancellations between the
coefficients of the effective Lagrangian of the
underlying model are unnatural. Even in this case, the resulting bounds
depend quite strongly on other parameters ($m_H$, $m_t$), and
anomalous couplings
of ${\cal O}(1)$ are still allowed by current data (see Section~2.4).

Present data from di-boson production at the Tevatron and from single
photon production at LEP yield bounds typically in the range of 0.5 --
3.0. They are summarized in Tables~1 and~5. $\Delta\kappa_\gamma$ is
currently constrained best by the process $p\bar p\to W^+W^-,\,
WZ\to\ell\nu jj$ (CDF), whereas the best bound on $\lambda_\gamma$
originates
from $W\gamma$ production at the Tevatron (D\O). The most precise
limits on the $Z\gamma V$ couplings result from
$e^+e^-\to\bar\nu\nu\gamma$ (L3; $h_3^Z$) and $Z\gamma$ production at
the Tevatron (D\O; $h_4^V$, $V=\gamma,\, Z$). Although the present
limits on $WWV$ and $Z\gamma V$ couplings are more than two orders of
magnitude larger than what one expects from theoretical considerations
if new physics exists at the TeV scale, these limits still provide
valuable information on how well the vector boson self-interaction
sector is tested experimentally at present.

\begin{figure}
\vskip 19cm
\includegraphics{fig16a.ps}
\includegraphics{fig16b.ps}
{\sl\noindent Figure~17: Comparison of the expected sensitivities on
anomalous $WWV$ couplings in the HISZ scenario from $e^+e^-\to
W^+W^-\to\ell\nu jj$ at LEP~II and various processes at the Tevatron.}
\end{figure}
Within the next 10~years, the limits on $WWV$ couplings are expected to
improve by more than one
order of magnitude by experiments conducted at the Tevatron and at
LEP~II. In Fig.~17 we compare the limits expected from
$e^+e^-\to W^+W^-\to\ell\nu jj$, $p\bar p\to
W^\pm\gamma\to e^\pm\nu\gamma$, $p\bar p\to W^\pm
Z\to\ell_1^\pm\nu_1\ell_2^+\ell_2^-$ and $p\bar p\to WW,\, WZ\to\ell\nu jj$,
$\ell^+\ell^- jj$ in the HISZ scenario [see Eqs.~(\ref{EQ:HISZ1})
--~(\ref{EQ:HISZ3})] for the envisioned energies and integrated
luminosities. The limits expected from future
Tevatron and LEP~II experiments for $\Delta\kappa_\gamma$ are quite
similar, whereas the Tevatron enjoys a clear advantage in constraining
$\lambda_\gamma$, if correlations between the two couplings are taken
into account. It should be noted, however, that the strategies to extract
information on vector boson self-interactions at the two machines are
very different. At the
Tevatron one exploits the strong increase of the anomalous
contributions to the helicity amplitudes with energy to derive limits.
At LEP~II, on the other hand, information is extracted from the angular
distributions of the final state fermions. Data from the Tevatron and
LEP~II thus yield complementary information on the nature of the $WWV$
couplings.

Because of the much higher energies accessible at the Tevatron and the
steep increase of the anomalous contributions to the helicity
amplitudes with energy, Tevatron experiments will be able to place
significantly better limits (of ${\cal O}(10^{-2}-10^{-3})$) on the
$Z\gamma V$ couplings than LEP~II ($\approx 0.5$). The Tevatron limits,
however, do depend non-negligibly on the form factor scale assumed.

At the LHC one expects to probe anomalous $WWV$ couplings with a
precision of ${\cal O}(10^{-1}-10^{-3})$ (see Table~4) if the form
factor scale
$\Lambda_{FF}$ is larger than about 2~TeV. Therefore, it may be
possible to probe anomalous $WWV$ couplings at the LHC at the level
where one would hope to see deviations from the SM. The limits on the
$Z\gamma V$ couplings are very sensitive to the value of
$\Lambda_{FF}$. For
$\Lambda_{FF}\ge 1.5$~TeV, the bounds which can be achieved are of ${\cal
O}(10^{-3})$ for $h_3^V$, and of ${\cal O}(10^{-5})$ for $h_4^V$. At the
NLC, $WWV$ and $Z\gamma V$ couplings can be tested with a precision of
$10^{-3}$ or better. Details depend quite sensitively on the center of
mass energy and the integrated luminosity of the NLC. If new physics
exists at the TeV scale, the NLC has the best chance to observe
deviations from the SM through anomalous $WWV$ couplings.

\vglue 0.3cm
{\elevenbf\noindent Acknowledgements}
\vglue 0.2cm

We would like to thank our colleagues in CDF, D\O, and L3 for many
valuable and stimulating discussions. We also thank Martin Einhorn for
his critical comments. This work has been supported in part by the
Department of Energy and by the N.~S.~E.~R.~C. of Canada and les Fonds
F.~C.~A.~R. du Qu\'ebec.

\vglue 0.3cm
{\elevenbf\noindent References}
\vglue 0.2cm

\end{document}